\documentclass[12pt]{article}
\usepackage{amsmath}
\usepackage{graphicx}
\usepackage{enumerate}
\usepackage{natbib}
\usepackage{url} % not crucial - just used below for the URL 

\allowdisplaybreaks
\newcommand{\blind}{0}

\usepackage{rotating}
\usepackage{color}

\usepackage{relsize}

\usepackage{multirow}
\def\E{\mathbb{E}}

\makeatletter
\newcommand*{\indep}{%
	\mathbin{%
		\mathpalette{\@indep}{}%
	}%
}

\newcommand*{\nindep}{%
	\mathbin{% % The final symbol is a binary math operator
		\mathpalette{\@indep}{\not}% \mathpalette helps for the adaptation
	}%
}
\newcommand*{\@indep}[2]{%
	\sbox0{$#1\perp\m@th$}% box 0 contains \perp symbol
	\sbox2{$#1=$}% box 2 for the height of =
	\sbox4{$#1\vcenter{}$}% box 4 for the height of the math axis
	\rlap{\copy0}% first \perp
	\dimen@=\dimexpr\ht2-\ht4-.2pt\relax
	\kern\dimen@
	{#2}%
	\kern\dimen@
	\copy0 % second \perp
}
\makeatother
 \usepackage{booktabs}
 \usepackage{amssymb}

\newtheorem{theorem}{Theorem}
\newtheorem{lemma}{Lemma}

\newtheorem{assumption}{Assumption}
\newtheorem{corollary}{Corollary}

\usepackage{caption}
\usepackage{subcaption}
\usepackage{amsfonts}
\def\E{\mathcal{E}}

\usepackage{subcaption}  % 添加subcaption包
\usepackage[all]{xy}
\usepackage{bm}
\usepackage{pifont}
\usepackage{stmaryrd}
\usepackage{pgf,tikz}
\usepackage{tikz}
\usetikzlibrary{shapes,arrows,positioning,calc}

\usepackage[flushleft]{threeparttable}
\usepackage{enumitem}
\usepackage{mathrsfs}

\def\pr{\textnormal{pr}}

\usepackage{arydshln}

\def\pr{\textnormal{pr}}

\usepackage{hyperref}
\hypersetup{
    colorlinks=true,
    linkcolor=blue,
    urlcolor=blue,
    citecolor=blue
}
\makeatletter

\usepackage{hyperref}

\usepackage[symbol]{footmisc}
 
\def\Pa{\mathrm{Pa}}

\usepackage{xr}
\makeatletter
\newcommand*{\addFileDependency}[1]{
  \typeout{(#1)}
  \@addtofilelist{#1}
  \IfFileExists{#1}{}{\typeout{No file #1.}}
}
\makeatother

\begin{document}

	\def\spacingset#1{\renewcommand{\baselinestretch}%
		{#1}\small\normalsize} \spacingset{1}

	\newcounter{savecntr}% Save footnote counter
\newcounter{restorecntr}
	%%%%%%%%%%%%%%%%%%%%%%%%%%%%%%%%%%%%%%%%%%%%%%%%%%%%%%%%%%%%%%%%%%%%%%%%%%%%%%
		\if0\blind
	{ 

    	\title{\bf  Assessing the Causes of Continuous Effects by Posterior Effects of Causes}
        
 			\author{ Shanshan Luo\textsuperscript{1},Yixuan Yu\textsuperscript{1},  Chunchen Liu\textsuperscript{2}, Feng Xie\textsuperscript{1}\thanks{Corresponding author.},  and      Zhi Geng\textsuperscript{1}     \\\\
			\textsuperscript{1} School of Mathematics and Statistics, \\Beijing Technology and Business University \\\\
  	\textsuperscript{2}  LingYang Co.Ltd, Alibaba Group, China}

		\date{}
		
		\maketitle
	} \fi
	
	\if1\blind
	{
		\bigskip
		\bigskip
		\bigskip
		\begin{center}
			{\LARGE\bf Title}
		\end{center}
		\medskip
	} \fi
		\medskip

\begin{abstract}  
To evaluate a single cause of a binary effect, \citet{dawid2014fitting} defined the probability of causation, while \citet{pearl2015causes} defined the probabilities of necessity and sufficiency.
For assessing the multiple correlated causes of a binary effect, \citet{lu2023evaluating} defined the posterior causal effects based on post-treatment variables. 
  In many scenarios, outcomes are continuous,   simply binarizing them and applying previous methods may result in information loss or biased conclusions. To address this limitation, we propose a series of posterior causal estimands for retrospectively evaluating multiple correlated causes  from a continuous effect, including  posterior intervention effects, posterior total causal effects, and  posterior natural direct   effects. 
Under the assumptions of sequential ignorability,    monotonicity, and perfect positive rank, we show that   the posterior causal estimands of interest are identifiable and present  the corresponding identification equations. We also provide a simple but effective estimation procedure and establish asymptotic properties of the proposed estimators.   An artificial hypertension example and a real developmental toxicity dataset are employed to illustrate our method.
 
\end{abstract} 
 \bigskip
 \bigskip
 \bigskip
\noindent%
\vfill

\newpage
\spacingset{1.9} % DON'T change the spacing!

\section{Introduction}
Causal inference includes not only assessing the effects of the causes but also deducing the causes from certain effects. \citet{pearl2000} introduced a three-level hierarchical structure known as the causal ladder: association, intervention, and counterfactual. The first two levels primarily involve prediction based on correlations or evaluating the effects of interventions.  In contrast, the third level focuses on assessing whether observed outcomes can be attributed to previous interventions or exposures, which is known as attribution analysis or assessing the causes of effects \citep{dawid2014fitting,dawid2015causes}.  For example, when considering an individual with a history of physical inactivity and hypertension, the first level involves predicting whether a new patient will develop hypertension. Subsequently, the second level involves assessing the causal effect of physical inactivity on blood pressure. Finally, the third level involves retrospectively examining whether his hypertension was caused by the lack of exercise. To address such retrospective questions, we need to imagine a counterfactual scenario in which he did exercise regularly and then infer his blood pressure at that time.   While randomized experiments and standard assumptions can effectively address the first two levels of the problem, they are insufficient for addressing the third level, posing challenges to traditional causal inference methods in solving such retrospective analysis problems.  %Randomized experiments and traditional assumptions used to identify causal effects are insufficient for detecting the causes of effects. 
The problem of inferring the causes of effects arises in many applications, such as social science \citep{vanderweele2012sufficient}, health risk assessment \citep{khoury2004epidemiologic}, legal contexts \citep{sanders2021differential}, and explainable artificial intelligence \citep{galhotra2021explaining}.

Most existing causal inference methods primarily focus on evaluating the effects of causes, with only a few aimed at deducing the causes of effects. \cite{pearl1999probabilities} outlined three counterfactual definitions of causal relationships to capture the necessity or sufficiency of a cause for a given binary effect. Additionally, \cite{dawid2014fitting} introduced the probability of causation to infer the cause for a given binary effect.  When there are multiple potentially correlated causes, \cite{lu2023evaluating} and \cite{Li2023BKA} introduced posterior causal effects under observed post-treatment variables to retrospectively deduce causes from a single effect and multiple effects, respectively. They showed the identiﬁability of these posterior causal effects under sequential ignorability and monotonicity assumptions.   In many practical applications, the outcome variables of interest may be continuous, such as weight, blood pressure, and income. For instance, when observing a hypertensive patient who does not exercise and has a history of heart disease, a physician may wish to infer how much their blood pressure would change if the patient had exercised in the past. However, most existing literature primarily conducts attribution analysis for binary outcomes of effects  \citep{pearl1999probabilities,dawid2014fitting,lu2023evaluating,Li2023BKA}.  Hence, there remains a research gap concerning the evaluation of causes of the continuous outcome or effect.
% In addition to continuous outcome variables, there is also interest in exploring the interaction of multiple correlated causes on a specific outcome, known as synergistic effects.   Current attribution analysis typically focuses on assessing the total causal effect of specific causes on a given outcome. For example, when observing a hypertensive patient with a lack of exercise and   history of heart disease, we also aim to infer whether the patient's   high blood pressure is caused by the synergistic effects of exercise and heart disease. 

In this paper, we consider evaluating multiple correlated causes of a given continuous effect. Specifically,  
we define the posterior intervention effects, posterior total causal effects,  posterior natural direct effects, and posterior natural  indirect effects  based on an event of interest for the continuous outcome variable, along with other evidence. 
Interestingly, we find that under the commonly used assumptions of sequential ignorability and perfect positive rank   \citep{heckman1997making}, the individual treatment effects and posterior intervention causal effects are identifiable. 
Additionally, assuming monotonicity among  multiple causes \citep{lu2023evaluating,Li2023BKA}, we demonstrate the identifiability of other posterior causal estimands, and provide their identification equations. We also present simplified identification results for the proposed causal estimands under a known  directed acyclic graph.  We introduce a two-step estimation approach: first estimating the counterfactual mappings and individual treatment effects for each   unit, and then estimating other posterior causal estimands. Importantly, our estimation method avoids the ill-posed inverse problem associated with inverting a nonlinear functional. We illustrate how these posterior causal estimands can evaluate multiple causes of a continuous effect  with  an artificial hypertension example and a real developmental toxicity dataset. Proofs of all theoretical results are provided in the supplementary materials.
 
\section{Notation and definitions}\label{Sec-notation-definition}
To introduce the notation and definitions in this paper, we first consider scenarios involving a single cause and a single effect or outcome. Let a binary variable $X$ represent a potential cause, where $X = 1$ indicates its presence and $X = 0$ indicates its absence. Let $Y$ represent the observed outcome. Let $Y_{X=0}$ and $Y_{X=1}$ respectively denote the potential outcomes of $Y$ under the conditions $X = 0$ and $X = 1$. For instance, in Example 2 of Section \ref{sec:exmas}, $X = 1$ could signify mice being exposed to a high dose of a toxic agent, and $X = 0$ means exposed to a low dose. 
$Y_{X=1}$ and $Y_{X=0}$ represent the potential body weights of mice exposed to high and low doses of the toxic agent, respectively. In many practical scenarios, researchers collect continuous outcome variables. For instance, measurements such as weight, blood pressure and income  are often continuous, but $Y$ may fall within a specific interval of interest, denoted as $\mathcal{E}$. For instance,  in  Example 1 of Section \ref{sec:exmas}, the event $\mathcal{E}=\mathbb{I}(Y>140)$ often indicates the presence of hypertension; in Example 2 of Section \ref{sec:exmas}, the event $\mathcal{E}=\mathbb{I}(Y<27)$ typically represents an underweight or abnormally thin mice. Therefore, when provided with observed evidence $(X = x, \mathcal{E})$, where $\mathcal{E}$ may represent some event of interest, we will investigate how likely it is that $X$ is a cause of such an event $\mathcal{E}$.

To measure how necessary or sufficient \( X \) is as a cause of an observed binary outcome, \citet{pearl2000} defined the probability of necessary causation or sufficient causation. To measure how likely it is that \( X \) is a cause of an observed binary outcome,  \citet{dawid2014fitting} defined the probability of causation. 
Next, we consider the case with multiple causes \( X = (X_1, \ldots, X_p) \) and a single outcome \( Y \), where \( X \) is a binary vector of causes, and the causes may affect each other.  Without loss of generality, we assume that the causes are arranged in a topological order such that $X_l$ is not a cause of $X_k$ for $k<l$. For example, $X$ is a sequence of observations ordered in time, or $X$ consists of variables in a directed acyclic graph where $X_k$ is not a descendant of $X_l$ for $k<l$. For generic sets of variables $W$ and $U$, we use $W_u$ to denote the potential outcome of $W$ that would have resulted if $U$ were intervened to level $u$. In particular, if $W=\left(W_1, \ldots, W_s\right)$, then $W_u=\left\{\left(W_1\right)_u \ldots,(W_s)_u\right\}$. We make the consistency assumption that connects observed variables to potential outcomes, i.e., $W_u=W$ if $U=u$ \citep{pearl2015causes}. We further suppose the composition assumption holds in the sense that for any variable sets $W, V$ and $U, W_{v u}=W_v$ if $U_v=u$ \citep{pearl2015causes}. The consistency assumption can be viewed as a special case of the composition assumption if $V$ is empty.
To measure how likely $X_k$ is a cause of the continuous effect given observed evidence $( x,\E ) $, we   extend the concept of \emph{posterior total causal effect} defined by \citet{lu2023evaluating} as follows: 
\begin{align*} 
\operatorname{PostTCE}\left(X_k \Rightarrow Y \mid x, \E\right)=E\left(Y_{X_k=1}-Y_{X_k=0} \mid x, \E\right),
\end{align*} 
where we use   ``$x$" to represent ``$X=x$" for notational simplicity.  As advocated by \citet{lu2023evaluating}, a larger value of the posterior total causal effect indicates that the effect or outcome is more attributable to the cause $X_k$. In practice, the cause that produces the largest a posterior total causal effect is usually considered the highest risk factor.

Similar to the direct causal effect considered by \citet{pearl2000},   we define the posterior natural direct effect of $X_k$ on $Y $   given the observed evidence $(x,\mathcal{E})$, which quantifies the effect of $X_k$ on $Y$ not mediated through intermediate variables.  Let $A_k=\left(X_1, \ldots, X_{k-1}\right)$ and $D_k=\left(X_{k+1}, \ldots, X_p\right)$. Then $X=\left(A_k, X_k, D_k\right)$, and $x=\left(a_k, x_k, d_k\right)$ denotes a value of $X$. Given the evidence $( x, \E ) $,  the {\it   posterior  natural direct effect} of $X_k$ on $Y$  is defined as follows:   $$\operatorname{PostNDE} (X_k \Rightarrow Y \mid x, \E )= E\{Y_{X_k=1, D_k(a_k,0)}-Y_{X_k=0 } \mid x,\mathcal{E}\} ,$$
where $ D_k(a_k,0)$ is the potential outcome under $(A_k ,X_k )=( a_k, 0)$. 
Throughout this paper, we use $D_k(a_k,x_k)$ and $(D_{k})_{a_k,x_k}$ interchangeably in the nested potential outcomes.  The posterior natural direct effect describes the effect observed when each individual in the subpopulation $(x, \mathcal{E} )$ switches from $X_k=0$ to $X_k=1$, while keeping $D_k$ at its value when $X_k=0$. It is important to note that this definition includes the event $\mathcal{E}$ defined by the observed outcomes.

Parallel to the natural indirect effect considered by \citet{pearl2000},  we also define the \textit{posterior natural indirect effect} of $X_k$ on $Y$ given the evidence $(X=x,\mathcal{E})$ as follows: $$\operatorname{PostNIE} (X_k \Rightarrow Y \mid x, \E )= E\{Y_{X_k=1 }-Y_{X_k=1, D_k(a_k,0)} \mid x,\mathcal{E}\} .$$  
The posterior natural indirect effect quantifies, for each individual in the subpopulation $(x, \mathcal{E})$, the effect observed when $X_k$ is set to $X_k=1$, while all intermediate variables along the pathway from $X_k$ to $Y$ change from  state $D_k(a_k,1)$ to state $D_k(a_k,0)$. Through the definitions, we have that: $ \mathrm{PostTCE}  (X_k \Rightarrow Y \mid x, \E )= \mathrm{PostNDE}(X_k \Rightarrow Y \mid x, \E )+  \mathrm{PostNIE}(X_k \Rightarrow Y \mid x, \E ) .$
% \begin{equation*} 
%     \begin{aligned}
%  &
% \end{aligned} 
% \end{equation*}
% We refer to posterior total causal effect, direct causal effects, and indirect causal effects together as posterior causal pathway estimands, as they are defined based on pathways similar to those in mediation analysis \citep{pearl2000}. 

Given the observed evidence $( x, \E ) $, in addition to assessing the total, direct, and indirect effects of a particular cause $X_k$, we also need to consider assessing synergistic effects by jointly intervening with all possible causes in another state $X=x^\prime$. Therefore, the {\it posterior intervention causal effect} for another state $X = x^\prime$ is defined as follows: 
 \begin{gather}
\label{eq:def-post-tce-X}
 \mathrm{PostICE}( Y_{ x^ \prime}    \mid x, \E)=E\left(Y_{ x^ \prime} -Y \mid x, \E\right) .
\end{gather}   
We illustrate the   definition \eqref{eq:def-post-tce-X} using Example 1 from Section \ref{sec:exmas}, where $x=(1,1)$ denotes heart disease and no exercise, while $x^\prime  =(0,0)$ denotes no heart disease and exercise, and $\E=\mathbb{I}(Y>140)$ denotes observed hypertension. $ \mathrm{PostICE}( Y_{ x^ \prime}  \mid x, \E)$ quantifies the synergistic reduction in blood pressure due to exercise and no heart disease.
Furthermore,  for any given observed evidence $(x,\mathcal{E})$, we can also characterize the difference in blood pressure between two different interventions    $X =x^\prime$ and $X = x^\ast$ as: $\mathrm{PostICE}(Y_{x^\prime} \mid x, \mathcal{E}) - \mathrm{PostICE}(Y_{x^\ast} \mid x, \mathcal{E})$.

% $ \mathrm{PostICE}( Y_{ x^\prime} ,Y_{ x } \mid x, \E)\geq \mathrm{PostICE}( Y_{x^\ast},Y_{ x } \mid x, \E)$ holds, this suggests that the patient's hypertension is more likely to be caused directly by the synergistic effect of inactivity and heart disease than by inactivity or heart disease alone.

The {\it  individual treatment effect} for any  pair $(x^\prime, x^\ast)$ can be defined as $\mathrm{ITE}(x^\prime,x^\ast)= Y_{x^\prime}-Y_{x^\ast} $, representing the difference in potential outcomes for each individual under two different treatment conditions. Inferring individual treatment effects presents a fundamental challenge because we can only observe one potential outcome for each unit \citep{rosenbaum1983central}. 
% Posterior direct causal effect depends on the ordering of the variables; in particular, the above equality might not hold for $X_p$ if the same variable $X_p$ is arranged forward in a different order $\left(X_1^{\prime}, \ldots, X_k^{\prime}=X_p, \ldots, X_p^{\prime}\right)$, see PostTCE and PostCDE of $X_3$ in Table 8. 
\section{Identiﬁability of posterior causal estimands and required assumptions}

\subsection{Assumptions required for identiﬁability}
Define $W=(X, Y)$ and let $W_{r: s}$ denote a subvector $(W_r, W_{r+1}, \ldots, W_s)$ of $W$ for $r \leq s$. Let $w_{r: s}^*=(w_r^*, \ldots, w_s^*) \preceq w_{r: s}=(w_r, \ldots, w_s)$ denote that $w_i^* \leq$ $w_i$ for all $r \leq i \leq s$. To identify the proposed estimands, we make the following commonly used  assumptions in literature \citep{heckman1997making,pearl2000,lu2023evaluating,Li2023BKA}.

\begin{assumption}[Sequential ignorability]
\label{assump:noconfounding}
We consider the following assumptions:
\begin{itemize}[leftmargin=30pt]
 \item[(i)]   there is no confounding between $W_s$ and $W_{1: s-1}$, i.e., $(W_s)_{w_{1: s-1}} \indep W_{1: s-1}$ for all $w_{1: s-1}$ and $s=2, \ldots, p+1$; 
 \item[(ii)] the elements in $\{(W_s)_{w_{1: s-1}}\}_{s=1}^{p+1}$ are mutually independent for any given $w_{1: p }$.

\end{itemize}
    
\end{assumption}

% Assumption \ref{assump:noconfounding}(i) means that the potential outcomes of each variable are independent of its precedent variables arranged in the causal order. Under Assumption \ref{assump:noconfounding}(i), if $W_s$ has a nonparametric causal structural model $W_s=m_s\left(W_{1: s-1}, \epsilon_s\right)$ with an unknown function $m_s$ and an error variable $\epsilon_s \indep W_{1: s-1}$, then Assumption \ref{assump:noconfounding}(ii) holds, because Assumption \ref{assump:noconfounding}(i) implies that $\epsilon_s \indep \epsilon_{1: s-1}$ for $s=2, \ldots, p+q$, which further implies Assumption \ref{assump:noconfounding}(ii). Assumption \ref{assump:noconfounding} rules out unobserved confounders among variables in $X$ and $Y$. However, each variable $X_k$ may still confound the relationships between $Y$ and $X_l$ or between $X_l$ and $X_s$ for $k<l, s$. When there is a set $H$ of observed background variables that are not affected by $X$, the independence in Assumption \ref{assump:noconfounding} can be relaxed to those conditional on $H$. No unobserved confounding assumption is commonly made in causal analysis from complex systems, such as causal mediation analysis \citep{Imai2010Stasci} or causal inference from longitudinal data with time-dependent treatments and confounders \citep{robins2000marginal}. 
 The independence condition in Assumption \ref{assump:noconfounding} can be relaxed by introducing the observed background variables. To simplify the exposition, we omit the background variables or baseline covariates in this paper.  The Assumption \ref{assump:noconfounding}(i) implies that the potential outcome of each variable is independent of the prior variables in causal order. Under the Assumption \ref{assump:noconfounding}(i), if $W_s$ has a nonparametric causal structural model $W_s=m_s\left(W_{1: s-1}, \epsilon_s\right)$  with an unknown function $m_s\left(\cdot, \epsilon_s\right)$ and a error  variable %\feng{should be error term/disturbance term???} 
$\epsilon_s \indep W_{1: s-1}$,  then Assumption \ref{assump:noconfounding}(ii) holds naturally because Assumption \ref{assump:noconfounding}(i) implies that $\epsilon_s \indep \epsilon_{1: s-1}$ for $s=2, \ldots, p+q$, which further implies Assumption \ref{assump:noconfounding}(ii). Assumption \ref{assump:noconfounding} rules out unobserved confounders between  any two variables in $W$. However, each variable $X_k$ may still confound the relationship between $Y$ and $X_l$, or between $X_l$ and $X_s$, provided $k<l, s$. Assumption \ref{assump:noconfounding} is frequently employed in causal analyses of complex systems, including causal mediation analysis \citep{Imai2010Stasci} and causal inference from longitudinal data involving time-dependent treatments and confounders \citep{robins2000marginal}.

% In the case of a single cause variable $X$, \citet{pearl2000} introduced the monotonicity assumption $Y_{X=0} \leq Y_{X=1}$, a requirement for the identifiability of the probability of necessity. When dealing with multiple cause variables and a binary outcome $Y$, \citet{lu2023evaluating} and \citet{Li2023BKA}
%  considered the monotonicity among multiple causes and the outcome, denoted as $(W_{p+1})_{w_{1: p}^*} \leq (W_{p+1})_{w_{1:p}}$ whenever $w_{1: s-1}^* \preceq w_{1: s-1}$ or equivalently $Y_{X=x^*} \leqslant Y_{X=x} $ whenever $ x^* \preceq x$. However, when the outcome variable is continuous, such monotonic relationships may no longer be applicable. 
% In the context involving a single cause and a continuous outcome, \citet{heckman1997making} emphasized the necessity of perfect positive rank correlation to recover the joint distribution $(Y_{X=0}, Y_{X=1})$, a condition sufficient for ensuring the identification of the individual treatment effect. Perfect positive rank correlation pertains to the dependence between potential outcomes across different treatment assignments. For scenarios involving multiple causes $X=(X_1,\ldots,X_p)$ and a continuous outcome $Y$, a more general requirement is described in the following assumption.

%

\begin{assumption}[Monotonicity]
\label{assump:monotonicity}
$(W_s)_{w_{1: s-1}^*} \leq(W_s)_{w_{1: s-1}}$ whenever $w_{1: s-1}^* \preceq w_{1: s-1}$ holds for $s=2, \ldots, p$.
\end{assumption}

To identify the posterior causal estimands, \citet{lu2023evaluating} and \citet{Li2023BKA}  also introduce the same monotonicity assumption across multiple potentially correlated causes. Assumption \ref{assump:monotonicity}   implies that each cause has a non-negative effect on subsequent causes. This assumption is commonly expressed in epidemiology as ``no prevention", meaning that no individual is helped by exposure to a risk factor. The validity of monotonicity cannot be tested directly, but under Assumption \ref{assump:noconfounding}, the monotonicity  can be falsified by imposing testable restrictions on the observed distribution. For example,  for any $w_{1: s-1}^* \preceq w_{1: s-1}$, the following equality can be used to falsify the monotonicity assumption: $
\operatorname{pr}(W_s=1 \mid W_{1: s-1}=w_{1: s-1}^*) \leq \operatorname{pr}(W_s=1 \mid W_{1: s-1}=w_{1: s-1}) $.

% To identify the posterior causal estimands,  \citet{lu2023evaluating} and \citet{Li2023BKA} also introduced the same monotonicity among the  multiple potentially correlated causes.  Assumption \ref{assump:monotonicity} implies that each cause has a non-negative effect on its outcome variables for all individuals. This assumption is commonly expressed in epidemiology as ``no prevention", i.e., no individual is helped by exposure to a risk factor. The validity of monotonicity cannot be tested directly, but this assumption can be falsified in some cases by imposing testable restrictions on the probability distribution of the observed data.  Similar assumptions are often made in studies of imperfect compliance of treatment  assignment 
% \citep{Angrist1996JASA}.

% {\red binary In the case of a single cause $X$, \citet{pearl2000}    introduced the monotonicity assumption for outcome variable $Y$, that is, $Y_{X=0} \leq Y_{X=1}$, which is crucial for the identifiability of probability of necessary causation and sufficient causation.  When dealing with multiple causes, \citet{lu2023evaluating} and \citet{Li2023BKA} consider the monotonicity of multiple causes with respect to a single binary outcome and multiple binary outcomes, respectively. }

\begin{assumption}[{Perfect positive rank}] 
\label{assump:rank-correlation}We assume that $W_{p+1}=m_{p+1}\left(W_{1: p}, \epsilon_{p+1}\right)$, or equivalently $Y = m_{p+1}(X ,\epsilon_{p+1})$, satisfies where $\epsilon_{p+1}$ represents a scalar-valued error variable, and the unknown link function $m_{p+1}(X,\cdot) $ is continuous and strictly increasing in $\epsilon_{p+1}$.
\end{assumption} 

For a given binary outcome, \citet{pearl2000} introduced the monotonicity assumption of the outcome variable with respect to a single causal variable $X$, denoted as $Y_{X=0} \leq Y_{X=1}$, which is crucial for identifying the probability of necessity or sufficiency under a single cause $X$. \citet{lu2023evaluating} considered the monotonicity assumption of the outcome variable with respect to multiple correlated causes $X$, denoted as $Y_{X=x^*} \leqslant Y_{X=x}$ for any $x^* \preceq x$. However, this monotonic relationship may not be applicable when the outcome variable is continuous. Therefore, we introduce Assumption \ref{assump:rank-correlation}, which is commonly used to identify individual treatment effects or quantile treatment effects in counterfactual causal inference literature \citep{heckman1997making,chernozhukov2005nonparametric}, but it is a relatively novel assumption in attribution analysis.

  The basic restriction in Assumption \ref{assump:rank-correlation}   is also referred to as the rank preservation or rank invariance  \citep{heckman1997making,chernozhukov2005nonparametric,vuong2017counterfactual,feng2020estimation}. The strict monotonic increase of $\epsilon$ can be changed to strict monotonic decrease without affecting the subsequent discussion. However, for simplicity, we assume strict monotonic increase. For any given error  $\epsilon_{p+1}^\ast$ and $x\neq x^\prime$, Assumption \ref{assump:rank-correlation} requires that the relative rank or quantile of $Y_x\equiv m_{p+1}(x,\epsilon_{p+1}^\ast)$ be the same as that of $Y_{x^\prime}\equiv m_{p+1}(x^\prime,\epsilon_{p+1}^\ast)$.  A stronger version of Assumption \ref{assump:rank-correlation} assumes that the error term $\epsilon_{p+1}$ is additive, that is, $Y=m_{p+1}^*(X)+\epsilon_{p+1}$ for some real-valued function $m_{p+1}^*(\cdot)$. Moreover, the  heteroscedasticity model also satisfies Assumption \ref{assump:rank-correlation}: $ Y=m_{p+1}^*(X)+\sigma(X)\epsilon_{p+1} $,  for some real-valued function $m_{p+1}^*(X)$ and positive function $\sigma(X)$.

\subsection{Identiﬁability and identiﬁcation equations of posterior causal estimands}
Under Assumptions \ref{assump:noconfounding} and \ref{assump:rank-correlation}, we first consider the identifiability of individual treatment effects and posterior intervention causal effects. Let $\mathcal{S}_{Y_x}$ denote the support of $Y_x$, which can be identified by $\mathcal{S}_{Y\mid X=x}$ under Assumption \ref{assump:noconfounding}. The key to our identification strategy is to match the potential outcome $Y_{x} \equiv m_{p+1}(x ,\epsilon_{p+1} )$ with the potential outcome $Y_{x^\prime}\equiv m_{p+1}(x^\prime,\epsilon_{p+1} )$ through a mapping $\phi_{x\to x^\prime}(\cdot)$, such that $Y_{x^\prime}=\phi_{x\to x^\prime}\left(Y_{  x}\right)$. This mapping $\phi_{x\to x^\prime}(\cdot)$ is termed a counterfactual mapping \citep{vuong2017counterfactual,feng2020estimation}, because it allows us to find the counterfactual outcome $Y_{x^\prime}$ from $Y_{x}$ using the function $\phi_{x\to x^\prime}(\cdot)$, and vice versa. Let $m_{p+1}^{-1}(x, \cdot)$ be the inverse function of $m_{p+1}(x, \cdot)$. According to Assumption \ref{assump:rank-correlation}, for any observed evidence $(X,Y)=(x,y)$, we can uniquely represent the error term as $\epsilon_{p+1}=m_{p+1}^{-1}(x, y)$, although the specific form of $m_{p+1}^{-1}(x,y)$ is unknown. Therefore, $Y_{x^\prime}$ is uniquely defined by $\phi_{x\to x^\prime}(y) \equiv m_{p+1}\{x^\prime , m_{p+1}^{-1}(x, y)\}$ for each $y \in \mathcal{S}_{Y_x}$. Moreover, the counterfactual mapping $\phi_{x\to x^\prime}(\cdot)$ is a continuous and strictly increasing function from $\mathcal{S}_{Y_x}$ onto $\mathcal{S}_{Y_{x^\prime}}$. Thus, if we can identify $\phi_{x\to x^\prime} (y)$ for all $y \in \mathcal{S}_{Y_x}$ and $x\neq x^\prime$, we can recover the individual treatment effects for every individual in the population. 
\begin{lemma}
\label{lem:ice}
  Under Assumptions \ref{assump:noconfounding} and   \ref{assump:rank-correlation}, for any $ y
\in \mathcal{S}_{Y_x}$,   the counterfactual mapping  $\phi_{x\to x^\prime}(\cdot)$  is identiﬁed   by the continuous extension of 
  \begin{equation}
      \label{eq:counter-mapping}
      \phi_{x\to x^\prime}(y)=F^{-1}_{ x^{\prime}}\{F _{ x}(y  )\},~~~ \forall y\in \mathcal{S}_{Y_x}^\circ,  
  \end{equation}
  where $F_x(y)=\pr(Y\leq y\mid X=x)$ and $\mathcal{S}_{Y_x}^\circ$ is the interior of $\mathcal{S}_{Y_x}$. 
 %for each unit $i$ with $X_i=x$ and $Y_i=y$,  the joint distribution of $2^p$ potential outcomes  is identifiable by  $ \{y_{i, X=(0\ldots0)}  ,\ldots,y_{i,X=(1\ldots1)}  \}$, where  $  y_{i,X=x^\prime}=F^{-1}_{X=x^{\prime}}\{F _{X=x}(y  )\}$ for $x\neq x^{\prime}$ and $ y_{i,X=x^\prime}=y$. 
  Moreover,  the individual treatment effects of every individual in the population can be identified.  
% The posterior controlled direct effect is also identified through the following equation:
%  \begin{align*} 
%  \operatorname{PostCDE} (X_k \Rightarrow Y_{D_k=d_k^*} \mid x, \E )=E(Y_{ x^{\prime}_1}-Y_{ x^{\prime}_0} \mid  x, Y>140) .
% \end{align*}
% where 
    
\end{lemma} 
Lemma \ref{lem:ice} establishes the identifiability of the counterfactual mapping $\phi_{x\to x^\prime}(\cdot)$ on $\mathcal{S}_{Y_x}$ in a constructive manner by matching the quantiles of $Y_{x}$ and $Y_{x'}$. We provide further intuitive explanation of \eqref{eq:counter-mapping} using Example 1 of Section \ref{sec:exmas}. If a patient has the highest blood pressure under exercise and no heart disease $x=(0,0)$ in the observed population $\{(X_i,Y_i):X_i=x\}$, then according to Assumption \ref{assump:rank-correlation}, their blood pressure under no exercise and heart disease $x^\prime=(1,1)$ should also be the highest in the observed population $\{(X_i,Y_i):X_i=x^\prime\}$; and vice versa. Thus,  we can recover any other potential outcome $Y_{x^\ast}$ of this patient by finding the highest blood pressure in the observed population $\{(X_i,Y_i):X_i=x^\ast\}$. Given the identifiability of the counterfactual mapping and individual treatment effects, the posterior intervention causal effect can also be identified using an inverse probability weighting expression  \citep{horvitz1952generalization}: 
%Lemma \ref{lem:ice} establishes the identification of the counterfactual mapping $\phi_x$ on $\mathcal{S}_{Y_x } $  in a constructive way by matching the quantiles of $Y_{ x}$ and  $Y_{ x}$. Given the recoverability of the  individual treatment effects, we can  also identify their  distributions. The posterior intervention causal effect can also be identified using an inverse probability weighting expression  as follows \citep{horvitz1952generalization}:
 \begin{align} 
 \label{eq:ipw-exp}
\mathrm{PostICE}\left( Y_{ x^{\prime}}  \mid  x, \E\right)=E\left[ \dfrac{\mathbb{I}(X=x, \E)}{\pr (X=x, \E)} \left\{ \phi_{x\to x^\prime}(Y )-Y \right\} \right],
\end{align} 
where $\mathbb{I}(\cdot)$ denotes the indicator function.

Assumptions \ref{assump:noconfounding} and \ref{assump:rank-correlation} establish the identifiability of individual treatment effects and  posterior intervention causal effects. 
However, they are not sufficient to ensure the identifiability of other posterior causal estimands (e.g., posterior natural direct causal effect) when considering a specific cause $X_k$. This is due to the challenge posed by identifying expectations of the nested potential outcomes given the observed evidence.
For example, given the evidence $X=(a_k, 1,d_k)$, the posterior natural direct causal effect involves the expectation of the cross-world intervention potential outcome $Y_{X_k=1,D_k(a_k,0)}$, which cannot be identified using Lemma \ref{lem:ice} alone. Before formally identifying these expectations, we provide another lemma for identifying the conditional probability of the counterfactual outcome of the causes.

% in combination with the   Assumption \ref{assump:noconfounding}(ii) and Assumption \ref{assump:monotonicity}, leading to
% \begin{align*}
% E\left\{Y_{ X_k=1,D_k(a_k,0)} \mid x,\E\right\}=\textstyle\sum_{d_k\preceq d_k^\ast} E(Y_{ a_k, 0,d_k^\ast} \mid x,\E)\pr\left\{D_k(a_k,0) =d_k^\ast\mid x  \right\}.
% \end{align*}
% The first term in the above expression is ensured by Lemma \ref{lem:ice}. However, the second term, $\pr \{D_k(a_k,0)=d_k^\ast \mid x \}$, essentially involves identifying the conditional probability of the counterfactual outcome. We now present the identifiability of the conditional probability $\pr \{D_k(a_k,0)=d_k^\ast \mid x \}$ under an additional Assumption \ref{assump:monotonicity}.
 
\begin{lemma}
\label{Lemma:monotonicity} 
Under Assumptions \ref{assump:noconfounding} and \ref{assump:monotonicity}, given the observed evidence $(a_k,x_k,d_k,\E)$, let $d_k^\ast=(x_{k+1}^\ast,\ldots,x_p^\ast)$ and $d_k =(x_{k+1} ,\ldots,x_p )$. 
\begin{itemize} 
    \item[(i)]For  $d_k^\ast\preceq d_k$, we have,
    \begin{align*}
         \pr \{D_k(a_k,0  )=d_k^\ast \mid a_k,1,d_k\}=\textstyle\prod_{s=k+1}^{p} \{    (1-x_s^\ast) +(2x_s^\ast-1) x_sR_{0s}\}  ,
    \end{align*}  
    where $   R_{0s}=  { \pr(X_{s} =1\mid a_k,0,x^\ast_{k+1 } ,\ldots,x^\ast_{s-1} )}/{ \pr (X_{s}=1\mid a_k,1,x_{k+1 } ,\ldots,x_{s-1}  )}.$
   
    \item[(ii)] For  $d_k\preceq d_k^\ast$, we have, 
    \begin{align*}
         \pr&\{D_k(a_k,1  )=d_k^\ast \mid a_k,0,d_k\}=\textstyle\prod_{s=k+1}^{p} \{x_s^\ast +(1-2x_s^\ast) (1-x_s)R_{1s}\}  ,
    \end{align*} where  
   $  R_{1s}=  { \pr(X_{s} =0\mid a_k,1,x_{k+1 } ^\ast,\ldots,x_{s-1}^\ast)}/{ \pr (X_{s}=0\mid a_k,0,x_{k+1 } ,\ldots,x_{s-1} )} .$ 
    %    \item[(iii)] When  $d_k= d_k^\ast$, we have:
    % \begin{align*}
    %      \pr&\{D_k(a_k,1  )=d_k^\ast \mid a_k,0,d_k\}=\textstyle\prod_{s=k+1}^{p} \{1-x_s x_s^\ast+(-1)^{1-x_s^\ast}x_s R_{1s}\}  ,
    % \end{align*}  
    % where $ R_{1s}= { \pr(X_{s} =1\mid a_k,0,x_{k+1:s-1} ^\ast)}/{ \pr (X_{s}=1\mid a_k,1, x_{k+1:s-1}  )}$.
\end{itemize}   
%     where    $\varphi_s(x_s,x_s^\ast,x_k,x_k^\ast,a_k, x_{k+1 :s-1}, x_{k+1 :s-1} ^\ast ) = {\pr \{X_{s}(a_{k}, x_{k :s-1} ^\ast)=x_s^\ast\mid X_{s}(a_{k}, x_{k :s-1})=x_s \}} $.  
%     \begin{gather*} 
%         \varphi_s(x_s,x_s^\ast,0,1;a_k, x_{k+1 :s-1}, x_{k+1 :s-1} ^\ast )  =1-(1-x_s) (1-x_s^\ast)+(-1)^{ x_s^\ast}(1-x_s) R_{0s},\\
%               \varphi_s(x_s,x_s^\ast,1,0;a_k, x_{k+1 :s-1}, x_{k+1 :s-1} ^\ast )  =1-x_s x_s^\ast+(-1)^{1-x_s^\ast}x_s R_{1s},
%             \\    \varphi_s(x_s,x_s^\ast,1,1;a_k, x_{k+1 :s-1}, x_{k+1 :s-1} ^\ast )  = \varphi_s(x_s,x_s^\ast,0,0;a_k, x_{k+1 :s-1}, x_{k+1 :s-1} ^\ast )  =\mathbb{I}(x_{k:s}=x_{k:s}^\ast) ,\\
%               R_{0s}=\dfrac{ \pr(X_{s} =0\mid a_k,0,x_{k+1:s-1} ^\ast)}{ \pr (X_{s}=0\mid a_k,1, x_{k+1:s-1}  )} ,~R_{1s}=\dfrac{ \pr(X_{s} =1\mid a_k,0,x_{k+1:s-1} ^\ast)}{ \pr (X_{s}=1\mid a_k,1, x_{k+1:s-1}  )}.
%     \end{gather*} 
 \end{lemma} 

\begin{theorem}
\label{thm:medeff}
Under Assumptions \ref{assump:noconfounding} and \ref{assump:monotonicity}, given the observed evidence $(a_k,x_k,d_k,\E)$, the posterior natural direct effect, posterior natural indirect effect, and  posterior total causal effect of $X_k$ on $Y$ can be identified using the following equations:
\begin{itemize}
\item[(i)] given $x_k=1$,  for any $x_k^\star\in\{0,1\}$, we have,
\begin{gather*}
E\{Y_{ x_k^\star,D_k(a_k,1)} \mid x,\E\} = E(Y_{x_k^\star,d_k } \mid x,\E),\\
E\{Y_{ x_k^\star,D_k(a_k,0)} \mid x,\E\} = \textstyle\sum_{d_k^\ast\preceq d_k} E(Y_{x_k^\star,d_k^\ast} \mid x,\E) \pr\{D_k(a_k,0)=d_k^\ast \mid x\},
\end{gather*}
\item[(ii)] given $x_k=0$, for any $x_k^\star\in\{0,1\}$, we have,
\begin{gather*}
E\{Y_{ x_k^\star,D_k(a_k,0)} \mid x,\E\} = E(Y_{x_k^\star,d_k } \mid x,\E),\\
E\{Y_{x_k^\star,D_k(a_k,1)} \mid x,\E\} = \textstyle\sum_{d_k\preceq d_k^\ast} E(Y_{x_k^\star,d_k^\ast} \mid x,\E) \pr\{D_k(a_k,1)=d_k^\ast \mid x\},
\end{gather*}
\end{itemize}
where the conditional expectation $E(Y_{x_k^\star,d_k^\ast} \mid x,\E)$ can be identified by Lemma \ref{lem:ice},  and   the conditional probability  $\pr\{D_k(a_k,x_k^\prime)=d_k^\ast \mid x\}$ can be identified by Lemma \ref{Lemma:monotonicity} for any $x_k^\prime\in\{0,1\}$.
\end{theorem}
Theorem \ref{thm:medeff} illustrates that with the additional monotonicity assumption \ref{assump:monotonicity}, we can also identify posterior
natural direct effect  and posterior total causal effect. 
In some cases, we may only want to conduct attribution analysis for continuous effects based on evidence from a subset  of $(X,\mathcal{E})$, while Lemma \ref{lem:ice} and Theorem \ref{thm:medeff} are based on fully observed evidence $(x ,\E)$. For some $s<p$, let $(X^{\prime},\mathcal{E})=(X_{i_1}, \ldots, X_{i_s},\mathcal{E})$ denote a subset of $(X,\mathcal{E})$. We can summarize the remaining set $X\backslash X^\prime$ to obtain the expected results for the subset $(X^{\prime},\mathcal{E})$  based on Lemma \ref{lem:ice} and Theorem \ref{thm:medeff}.  To simplify the exposition, we omit this part. We also refer to Corollary 1 in \citet{lu2023evaluating} and Theorem 3 in \citet{Li2023BKA} for parallel results.
\subsection{Identification equations of posterior causal effects under causal networks}
 In this section, we consider the causal structure of observed variables $(X, Y)$ represented by a directed acyclic graph or network. We aim to present the simplified identification expressions for the posterior causal estimands given a known directed acyclic graph.   For $k=1, \ldots, p$, let $\Pa_k$ and $\Pa_Y$ denote the sets of parents of $X_k$ and $Y$ in the graph, respectively. The joint probability distribution $\operatorname{pr}(X, Y)$ can be factorized as $\operatorname{pr}(X, Y)=\prod_{k=1}^p \operatorname{pr}(X_k \mid \Pa_k)\operatorname{pr}(Y  \mid \Pa_Y)$. We assume $\operatorname{pr}(x, y)>0$ for each $(x, y)$. For a given graph, the sequential ignorability assumption (Assumption \ref{assump:noconfounding}) posits that there is no unobserved variable intervening between any two or more nodes in the graph. The monotonicity assumption \ref{assump:monotonicity} implies that each node $X_k$ has a positive individual monotonic effect on its child nodes. Furthermore, Assumption \ref{assump:rank-correlation} can be simplified such that the unknown link function $m_{p+1}(\mathrm{Pa}_Y,\epsilon_{p+1})$ only needs to be continuous with respect to $\mathrm{Pa}_Y$ and strictly increasing in $\epsilon_{p+1}$. Specifically, the continuous outcome variable is only required to perfectly match the quantiles of potential outcomes under different realizations of its parent nodes $\mathrm{Pa}_Y$. We define a simpler counterfactual mapping $\phi_{\mathrm{pa}_y\to \mathrm{pa}_y^\prime}(\cdot)$ to characterize the mapping relationship of the parent nodes of $Y$ under different realizations $\mathrm{pa}_y$ and $ \mathrm{pa}_y^\prime$. The theoretic results presented in Lemma \ref{lem:ice} and \eqref{eq:ipw-exp} can be simplified   for a given graph. 
  
\begin{corollary}
\label{coro:ice-dag}
Suppose that the causal network of $\left(X_1, \ldots, X_p, Y\right)$ is a directed acyclic graph. Then, under  Assumptions \ref{assump:noconfounding} and   \ref{assump:rank-correlation},  for any $x\neq x^\prime$ and $ y
\in \mathcal{S}_{Y_x}$,    let $\mathrm{pa}_y\subset x$ and $\mathrm{pa}_y^\prime\subset x^\prime$, 
  the counterfactual mapping  $\phi_{x\to x^\prime}(\cdot)$  can be reduced as follows:
  \begin{equation*} 
\phi_{x\to x^\prime}(y)=      \phi_{\mathrm{pa}_y \to\mathrm{pa}^\prime_y}(y)=F^{-1}_{\mathrm{pa}^\prime_y}\{F _{ \mathrm{pa}_y }(y  )\},~~~ \forall y\in \mathcal{S}_{Y_x}^\circ,  
  \end{equation*}
  where $m_{ \mathrm{pa}_y }(y)=\pr(Y\leq y\mid \mathrm{Pa}_Y=\mathrm{pa}_y )$. 
 %for each unit $i$ with $X_i=x$ and $Y_i=y$,  the joint distribution of $2^p$ potential outcomes  is identifiable by  $ \{y_{i, X=(0\ldots0)}  ,\ldots,y_{i,X=(1\ldots1)}  \}$, where  $  y_{i,X=x^\prime}=F^{-1}_{X=x^{\prime}}\{F _{X=x}(y  )\}$ for $x\neq x^{\prime}$ and $ y_{i,X=x^\prime}=y$. 
  Moreover,  the individual treatment effects  of every individual in the population can be identified.  The following equation also holds:
 \begin{align*} 
E\left(Y_{ x^{\prime}} \mid  x, \E\right)&=E(Y_{ \mathrm{pa}^{\prime}_y} \mid \mathrm{pa}_y, \E)  =E\left\{\frac{\mathbb{I}(\mathrm{Pa}_y = \mathrm{pa}_y, \E)}{\pr (\mathrm{Pa}_y = \mathrm{pa}_y, \E)} \phi_{\mathrm{pa}_y \to \mathrm{pa}^\prime_y}(Y )   \right\} ,
\end{align*}
and 
the posterior intervention causal effect is identifiable.
\end{corollary} 
Corollary \ref{coro:ice-dag} indicates that the counterfactual mapping depends only on $\Pa_Y$, thus the posterior intervention causal effect and individual treatment effects can be calculated using the low-dimensional conditional probabilities $\pr(Y\leq y\mid  \mathrm{pa}_y )$.  Similarly, the identification equations of the conditional probability $\pr \{D_k(a_k,0)=d_k^\ast \mid x \}$ can also be simplified by replacing $R_{0s}$ and $R_{1s}$ in Lemma \ref{Lemma:monotonicity}  with the following expressions:
 \begin{align}
 \label{eq:reduce-prob}
                   R_{0s}^\ast ={ \pr(X_{s} =0\mid \mathrm{pa}_s^\ast)}/{ \pr (X_{s}=0\mid  \mathrm{pa}_s  )} ,~~~R_{1s}^\ast ={ \pr(X_{s} =1\mid \mathrm{pa}_s^\ast)}/{ \pr (X_{s}=1\mid  \mathrm{pa}_s )},
 \end{align}
where $ \mathrm{pa}_s^\ast\subset( a_k,0,x^\ast_{k+1 } ,\ldots,x^\ast_{s-1})$ and $ \mathrm{pa}_s \subset( a_k,1,x _{k+1 } ,\ldots,x _{s-1} )$. 
The following Corollary \ref{coro:pa-reduce-thm}  provides a simplified identification equation of other posterior causal estimands for a given graph.

\begin{corollary}
\label{coro:pa-reduce-thm}
Suppose that the causal network of $\left(X_1, \ldots, X_p, Y\right)$ is a directed acyclic graph. Under Assumptions \ref{assump:noconfounding} and \ref{assump:monotonicity}, given the observed evidence $(a_k,x_k,d_k,\E)$, the posterior natural direct effect, posterior natural indirect effect, and posterior  total causal effect of $X_k$ on $Y$ can be identified using the following equations:
\begin{itemize}
\item[(i)] when $x_k=1$, for any $x_k^\star\in\{0,1\}$, we have,
\begin{gather*}
E\{Y_{ x_k^\star,D_k(a_k,1)} \mid x,\E\} = E(Y_{\mathrm{pa}^\star} \mid \mathrm{pa}_y,\E),\\
E\{Y_{ x_k^\star,D_k(a_k,0)} \mid x,\E\} = \textstyle\sum_{d_k^\ast\preceq d_k} E(Y_{\mathrm{pa}^\ast_y} \mid \mathrm{pa}_y,\E) \pr\{D_k(a_k,0)=d_k^\ast \mid x\},
\end{gather*}
\item[(ii)] when $x_k=0$, for any $x_k^\star\in\{0,1\}$, we have,
\begin{gather*}
E\{Y_{ x_k^\star,D_k(a_k,0)} \mid x,\E\} = E(Y_{\mathrm{pa}^\star} \mid \mathrm{pa}_y,\E),\\
E\{Y_{x_k^\star,D_k(a_k,1)} \mid x,\E\} = \textstyle\sum_{d_k\preceq d_k^\ast} E(Y_{\mathrm{pa}^\ast_y} \mid \mathrm{pa}_y,\E) \pr\{D_k(a_k,1)=d_k^\ast \mid x\},
\end{gather*}

\end{itemize}
 where  the conditional expectations  $E(Y_{\mathrm{pa}_y^\star} \mid \mathrm{pa}_y,\E)$ and $E(Y_{\mathrm{pa}_y^\ast} \mid \mathrm{pa}_y,\E) $ can be identified by Corollary \ref{coro:ice-dag}  for ${\mathrm{pa}}^\star_y\subset (a_k,x_k^\star,d_k)$  and  ${\mathrm{pa}}^\ast_y\subset (a_k,x_k^\star,d_k^\ast)$, and the conditional probability $\pr\{D_k(a_k,x_k^\prime)=d_k^\ast \mid x\}$ can be identified by Lemma \ref{Lemma:monotonicity} and \eqref{eq:reduce-prob} for any $x_k^\prime\in\{0,1\}$.
\end{corollary} 
\section{Estimation}
\label{sec:estimation}
In this section, we propose a simple but effective method for estimating the counterfactual outcome mapping $\phi_{x\to x^\prime}(\cdot )$ as well as the posterior causal estimands. Let $\{(X_i,Y_i ) : i=1, \cdots, n\}$ be the independent and identically distributed samples generated according to Assumptions \ref{assump:noconfounding}-\ref{assump:rank-correlation}. Our estimation procedure consists of two steps: first, for each observation $\left( X_i,Y_i \right)=(x,y)$, we estimate the counterfactual mapping $\phi_{x\to x^\prime}(y )$ for  $x^\prime\neq x$ using a simple estimator that minimizes a convex population objective function and constructs pseudo samples of the counterfactual outcomes for all individuals. In the second step, we nonparametrically estimate the posterior causal estimands based on the counterfactual mapping $\phi_{x\to x^\prime}(\cdot )$.    For simplicity, let $\mathcal{S}_{Y_{x}}$ denote a compact interval $[y_{x}^l, y_{x}^u]$ for any $X=x$, where $-\infty<y_{x}^l<y_{x}^u<+\infty$. We also assume the compact support $[y_{x}^l, y_{x}^u]$ is known. Otherwise, it can be estimated using the methods proposed in \citet{korostelev2012minimax}.   To establish the asymptotic properties of the estimators to be proposed in this section, we make the following assumptions.

\begin{assumption}
\label{assump:error}
     (i) The function $m_{p+1}(x,\epsilon_{p+1})$ is continuously differentiable in error term 
 $\epsilon_{p+1}$ for any $X=x$; (ii) The probability density function of the error term $ {\epsilon _{p+1} } $ is continuous; (iii)    $\inf _{y \in [y_{x}^l, y_{x}^u]} g_{ x}(y)>0$ for any $X=x$, where  ${  g_{x}(y)} = \partial F_{x} (y) /\partial  y$. 
\end{assumption} 
Assumptions \ref{assump:error}(i) and \ref{assump:error}(ii) are regularity conditions and, together with Assumptions  \ref{assump:noconfounding} and   \ref{assump:rank-correlation}, ensure that the marginal distribution $F_{x }(y)$ is absolutely continuous with respect to the Lebesgue measure, and its probability density function $g_{x  }(y)$ is also continuous for any $X=x $. 
Assumption \ref{assump:error}(iii) is introduced for the sake of simplicity in explanation. Trimming techniques can be employed to relax this assumption, but they may introduce technical complexities. 

We now aim to derive a counterfactual outcome mapping $\phi_{x\to x^\prime}(\cdot)$ from two marginal distributions $F_{x}(\cdot)$ and $F_{x ^\prime}(\cdot)$. For a given $y \in \mathbb{R} $, we define the objective function as follows:
\begin{align*} 
\rho_{x\to x^\prime}(t ; y )&= {E}\left\{\operatorname{sign}(Y-y) \mid X=x\right\} \times t-{E}(|Y-t| \mid X=x^\prime ), 
\end{align*}
where $\operatorname{sign}(u) \equiv 2 \times \mathbb{I}(u>0)-1$. The above objective function is motivated by the quantile
regression method in \citet{koenker1978regression} and \citet{feng2020estimation}. Simple calculations reveal that the first-order and second-order conditions of this objective function are given by:
\begin{align*} 
\frac{\partial \rho_{x\to x^\prime}(t ; y )}{\partial t}=2\left\{F_{x^\prime}(t)-F_{x}(y)\right\}  =0,~~ \frac{\partial^2 \rho_{x\to x^\prime}(t ; y )} {\partial t^2}=2   {  g_{ x^\prime}(t)} \geq 0.
\end{align*}   The following lemma demonstrates that for any $y \in \mathcal{S}_{Y _x}^{\circ}$, the objective function $\rho_{x\to x^\prime}(\cdot ; y )$ is uniquely minimized at the counterfactual outcome $\phi_{ x\to x^\prime}(y)$.

\begin{lemma}
\label{lem:min}
Under Assumptions \ref{assump:noconfounding} and \ref{assump:monotonicity}, $\rho_{x\to x^\prime}(\cdot ; y)$ is continuously differentiable and weakly convex on $\mathbb{R}$. Additionally, $\rho_{x\to x^\prime}(\cdot ; y)$ is strictly convex on $\mathcal{S}_{Y_{x^\prime}}^{\circ}$ and uniquely minimized on $\mathbb{R}$ at $\phi_{x\to x^\prime}(y)$ when $y \in \mathcal{S}_{Y_{x }}^{\circ}$.
\end{lemma} 
Lemma \ref{lem:min} provides the foundation for our nonparametric estimation of the counterfactual mappings and posterior causal estimands.  For the $i$ th observational unit $(X_i=x, Y_i)$, let
\begin{align*} 
\hat{\rho}_{x\to x^\prime}\left(t ; Y_i\right)=\dfrac{\sum_{j =1 }^n \operatorname{sign}(Y_j-Y_i) \times \mathbb{I} (  X_j=x )}{\sum_{j =1 }^n\mathbb{I} (X_j=x  )} \times t-\dfrac{\sum_{j =1 }^n\left|Y_j-t\right| \times \mathbb{I}(  X_j=x^{\prime} )}{\sum_{j =1 }^n \mathbb{I}(X_j=x^{\prime}  )}   .
\end{align*}   Hence, we can estimate the counterfactual outcome of the $i$ th   unit by,
\begin{align} 
\label{eq:coun-map}
\hat{\phi}_{x\to x^\prime}\left(Y_i\right)=\underset{t \in[y_{x^{\prime}}^l, y_{x^{\prime}}^u]}{\arg \min } \hat{\rho}_{x\to x^\prime}\left(t ; Y_i \right), & ~~\text { if } X_i=x  .
\end{align} 
 It is worth noting that the proposed method for estimating counterfactual mappings is computationally simple and avoids issues associated with the inverse problem.  Specifically, we can solve the one-dimensional optimization problem using grid search algorithm, which is both simple and robust. 
 The next theorem establishes the uniform consistency and $\sqrt{n}$-asymptotic distribution of the counterfactual mapping estimator $\hat{\phi}_{x\to x^\prime} (\cdot)$ over its full support. 

\begin{theorem}
\label{thm:emipircal-process}
     Suppose Assumptions  \ref{assump:noconfounding}-\ref{assump:error}   hold. Then, for any pair $ (x,x^\prime) $, we have,
\begin{align*}
\sup _{y \in[y_x^l, y_x^u ]}\left|\hat{\phi}_{x\to x^\prime}(y)-\phi_{x\to x^\prime}(y)\right|=o_p(1) .
\end{align*}
Moreover, the empirical process $g_{x ^\prime}\left\{\phi_{x\to x^\prime}(\cdot)\right\} \times \sqrt{n}\left\{\hat{\phi}_{x\to x^\prime}(\cdot)-\phi_{x\to x^\prime}(\cdot)\right\}$ converges in distribution to a zero-mean Gaussian process with covariance kernel $\Sigma_{ x}(s,t)$ for any $s\leq t$:
% \begin{align*}
%   \dfrac{F_x(y)}{\pr(X=x)}+\dfrac{F_{x^\prime}\{{\phi}_{x\to x^\prime}(y)\}}{\pr(X=x^\prime)} -\left[    {F_{x^\prime}\left\{{\phi}_{x\to x^\prime}(y)\right\}} - {F_{x }(y) }  \right]\left[    {F_{x^\prime}\left\{{\phi}_{x\to x^\prime}(y^\ast)\right\}} -{F_{x }(y^\ast)  }   \right] . 
% \end{align*}  
\begin{align*}
     \dfrac{F_{x^\prime}\left\{{\phi}_{x\to x^\prime}(s)\right\} \left[1-F_{x^\prime}\left\{{\phi}_{x\to x^\prime}(t)\right\}\right]}{\pr(X=x^\prime)}  +\dfrac{F_{x } (s) \{1-F_{x } (t) \}}{\pr(X=x )} 
\end{align*}
\end{theorem}
Theorem \ref{thm:emipircal-process} reveals the uniform convergence of $\hat{\phi}_{x\to x^\prime}(\cdot)$, and Assumption \ref{assump:error} ensures that the boundary is also included within it. When $ s=t=y$,   the asymptotic variance of $g_{x^\prime}\left\{\phi_{x\to x^\prime}(y)\right\} \times \sqrt{n} \{\hat{\phi}_{x\to x^\prime}(y)-{\phi}_{x\to x^\prime}(y)\}$ can be expressed as follows:
\begin{align*} 
\Sigma_{ x}^2(y) \equiv    \dfrac{F_{x^\prime}\left\{{\phi}_{x\to x^\prime}(y)\right\} \left[1-F_{x^\prime}\left\{{\phi}_{x\to x^\prime}(y)\right\}\right]}{\pr(X=x^\prime)}  +\dfrac{F_{x } (y) \{1-F_{x } (y) \}}{\pr(X=x )} .
\end{align*} 
 We note that the asymptotic variance of $\hat{\phi}_{x\to x^\prime}(y)$ is inversely related to $g_{x^\prime }^2\left\{\phi_{x\to x^\prime}(y)\right\} $, but is independent of the magnitude of the individual treatment effect. In addition, the proportions of the subpopulations $X=x$ and $X=x^\prime$ have a significant effect on the variance, with smaller proportions resulting in larger asymptotic variance; indeed these two proportions reflect the effective sample size available to practitioners. Finally, we find that the asymptotic variance decreases to zero as $y$ approaches its boundary. Thus, we get more accurate estimates when the counterfactual outcomes are closer to the boundary.
 
%  It is worth pointing out that our estimation of counterfactual mapping is computationally simple and does not suffer from an ill-posed inverse problem. In particular, to solve the one-dimensional optimization problem, the practitioner can use a grid search algorithm that is simple but highly robust. As a matter of fact, the sample objective function $\hat{\rho}_{x\to x^\prime}\left(y_i\right)$ is piecewise linearly continuous and minimized at some observation(y) $Y_j$ of the dependent variable where $X_j=x$. Moreover, by minimizing either $\hat{\rho}_{x^\prime}\left(y_i\right)$   for each observation $i$ in the sample, respectively, we obtain a pseudo sample of ITEs as follows: $ \mathbb{I}(X_i=x) \times\left\{Y_i-\hat{\phi}_{x\to x^\prime}\left(Y_i\right)\right\}+\mathbb{I}(X_i=x^\prime) \times\left\{\hat{\phi}_{x\to x^\prime}\left(Y_i\right)-Y_i\right\}$. 
% Given the uniform $\sqrt{n}$-consistency of $\hat{\phi}_{x\to x^\prime}$, it follows that $\hat{\Delta}_i$ also uniformly converges to $\Delta_i$ at the $\sqrt{n}$-rate. 
% In addition, there will also be $\sqrt{n}$-rate consistency about other posetrior causal estimands. 

By separately minimizing $\hat{\rho}_{x\to x^\prime}\left(\cdot ; Y_i\right)$ for each unit $i$ with $(X_i=x, Y_i)$, we can estimate the counterfactual outcome for an individual $i$ under state $X= x^\prime$ by the counterfactual mapping $\hat{\phi}_{x\to x^\prime} \left(Y_i \right)$   in \eqref{eq:coun-map}, and the individual treatment effect $\mathrm{ITE}(x^\prime,x^\ast)$ can be estimated as follows: 
$$ \hat{\Delta}_{\scriptstyle{\mathrm{ITE},{i}}}(x^\prime,x^\ast)=\hat{\phi}_{x\to x^\prime}\left(Y_i\right)-\hat{\phi}_{x\to x^\ast}\left(Y_i\right). $$
Moreover, other posterior causal estimands can also be constructed using similar moment estimators. For instance, the posterior intervention causal effect in \eqref{eq:ipw-exp} can be estimated as follows:
$$ \hat\Delta_{\scriptstyle\mathrm{PostICE}}\left( Y_{ x^{\prime}}  \mid  x, \E\right)=\textstyle \sum_{i=1}^n {\mathbb{I}(X_i=x, \E_i)}  \left\{ \hat\phi_{x\to x^\prime}(Y_i )- Y_i \right\}\big/  \textstyle  \sum_{i=1}^n {\mathbb{I}(X_i=x, \E_i)}, $$
where $\E_i$ denotes the event that individual $i$ experiences the event of interest $\E$.
Given the $\sqrt{n}$-consistency of $\hat{\phi}_{x\to x^\prime}(Y_i)$ and $\hat{\phi}_{x\to x^\ast}(Y_i)$ in Theorem \ref{thm:emipircal-process}, we know that the estimators $ \hat{\Delta}_{\scriptstyle{\mathrm{ITE},{i}}}(x^\prime,x^\ast)$ as well as $\hat\Delta_{\scriptstyle\mathrm{PostICE}}\left( Y_{ x^{\prime}}  \mid  x, \E\right)$ also converge uniformly at the $\sqrt{n}$-rate \citep{van1996weak}.
\section{Examples}
\label{sec:exmas}
\subsection{Example 1: Risk factors for hypertension}
\begin{figure}[t]
    \centering
    \includegraphics[width=\textwidth]{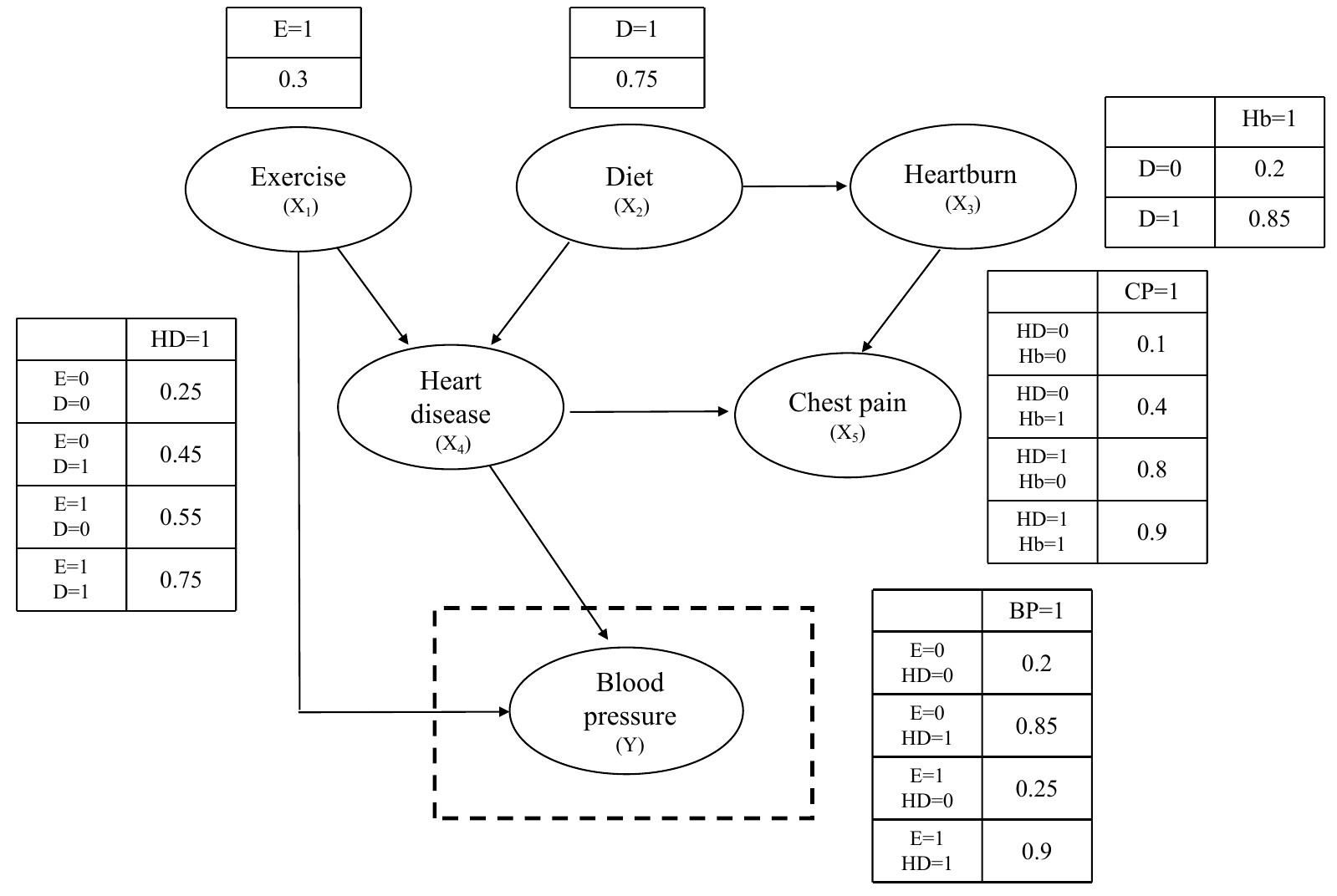}
    \caption{A causal network representing hypertension and its risk factors, where $\mathrm{U}(a,b)$ denotes a uniform distribution on the interval $[a,b]$.}
    \label{fig:BP-DAG}
\end{figure}

In this section, we assess the posterior causal effects of risk factors on hypertension utilizing the example provided in \citet{lu2023evaluating}.  
Figure \ref{fig:BP-DAG} presents the causal network and the corresponding conditional probabilities, where Exercise (E), Diet (D), Heart Disease (HD), Heartburn (Hb), and Chest Pain (CP) are potential causes of hypertension. Let $\mathrm{E}=1$ denote no daily exercise, $\mathrm{D}=1$ denote unhealthy diet, $\mathrm{HD}=1$ denote heart disease, $\mathrm{Hb}=1$ denote heartburn, $\mathrm{CP}=1$ denote chest pain, and $\mathrm{BP}$ denote continuous blood pressure $Y$.  
When considering continuous outcome variable, we regenerated the data without unobserved confounders to satisfy Assumption \ref{assump:noconfounding}. We also ensured that the relationships between the causes satisfied the monotonicity assumption (Assumption \ref{assump:monotonicity}). Additionally, the various potential outcomes satisfy the perfect rank   assumption (Assumption \ref{assump:rank-correlation}). Specifically, we modeled blood pressure $Y$ as a continuous outcome variable and ensured that, after binarization, its distribution matches that observed in \citet{lu2023evaluating}. 
  See supplementary materials for more details. For instance, lack of daily exercise $(\mathrm{E}=1)$ and poor diet $(\mathrm{D}=1)$ are not preventive for heart disease $\mathrm{HD}$. According to the causal network, the topological order of variables is $X=\left(X_1, \ldots, X_5\right)=(\mathrm{E}, \mathrm{D}, \mathrm{Hb}, \mathrm{HD}, \mathrm{CP})$.  The joint probability of $X$ and $Y$ is calculated by substituting the conditional probabilities from Figure \ref{fig:BP-DAG}  into the following equation: 
$ 
\operatorname{pr}(\mathrm{E}, \mathrm{D}, \mathrm{Hb}, \mathrm{HD}, \mathrm{CP}, \mathrm{BP})=\operatorname{pr}(\mathrm{E}) \operatorname{pr}(\mathrm{D}) \operatorname{pr}(\mathrm{Hb} \mid D) \operatorname{pr}(\mathrm{HD} \mid \mathrm{E}, \mathrm{D}) \operatorname{pr}(\mathrm{CP} \mid \mathrm{Hb}, \mathrm{HD}) \operatorname{pr}(\mathrm{BP} \mid \mathrm{HD}).
$  Without loss of generality, we consider the event $\E=\mathbb{I}(Y>140)$, which indicates whether hypertension is present. 
The posterior causal estimands can be used to analyze the causes of hypertension for a patient based on the evidence $(x, Y>140)$.

\begin{table}[t]
\centering
\caption{Results of posterior intervention causal effects based on
different evidences.}
\label{tab: NLSY-res}   
\resizebox{0.948\columnwidth}{!}{% 
\begin{tabular}{ccccccc}
 \toprule   \addlinespace[0.5mm]\addlinespace[0.25mm]
    $ \mathrm{PostICE} (Y_{ x^{\prime}}   \mid  x, Y>140)$   && $(x_1,x_4)=(0,0)$ & $(x_1,x_4)=(0,1)$ & $(x_1,x_4)=(1,0)$  &$(x_1,x_4)=(1,1)$  \\ 
 \addlinespace[0.25mm]\cline{3-6} \addlinespace[1mm] 
 $(x_1^\prime,x_4^\prime)=(0,0)$ &&0.00 & -12.43 & -1.76 & -18.19 \\ \addlinespace[0.25mm]
 $(x_1^\prime,x_4^\prime)=(0,1)$ & &  3.34 & 0.00 & 2.34 & -5.43 \\ \addlinespace[0.25mm]
 $(x_1^\prime,x_4^\prime)=(1,0)$ & &1.75 & -10.01 & 0.00 & -15.75 \\ \addlinespace[0.25mm]
 $(x_1^\prime,x_4^\prime)=(1,1)$& &  12.05 & 5.57 & 10.86 & 0.00 \\ \addlinespace[0.25mm]\bottomrule 
\end{tabular} }
\end{table}
% We first present the posterior intervention causal effects  for hypertension $\E=\mathbb{I}(Y>140)$  based on different observed evidence, as shown in Table \ref{tab: NLSY-res}. According to Corollary \ref{coro:ice-dag}, it is known that the  postICEs are only related to the parent nodes of  BP, namely \(X_1\) and \(X_4\), representing exercise   and heart disease, respectively. We find that in the subgroup $\left(x_1, x_4,\E\right)=(1,1,Y > 140)$, the largest posterior intervention causal effect will occur. Specifically, for individuals who do not exercise, have heart disease, and have high blood pressure, if they had exercised previously and did not have heart disease, i.e., receiving treatment \((x_1^\prime, x_4^\prime)=(0,0)\), their blood pressure would significantly decrease by 18.19. 
% Conversely, in the population where $\left(x_1, x_4,\E\right)=(0,0,Y > 140)$, i.e., individuals who exercise, do not have heart disease, and have high blood pressure, if they had not exercised previously and did not have heart disease, i.e., receiving treatment \((x_1, x_4)=(1,0)\), their blood pressure would slightly increase by 1.75.

We first present the posterior intervention causal effects for hypertension,   based on different observed evidence, as shown in Table \ref{tab: NLSY-res}. According to Corollary \ref{coro:ice-dag}, we know that the post intervention causal effects are only related to the parent nodes of BP, namely \(X_1\) (exercise) and \(X_4\) (heart disease). We find that the largest change in posterior intervention causal effect occurs with observed evidence $\left(x_1, x_4, \E\right) = (1,1, Y > 140)$ and the intervened treatment $(x_1^*, x_4^* ) = (0,0)$. 
 Specifically, for individuals who do not exercise, have heart disease, and have high blood pressure, if they had exercised previously and did not have heart disease, i.e., receiving treatment \((x_1^\prime, x_4^\prime) = (0,0)\), their blood pressure would significantly decrease by 18.19. Conversely, in the population where $\left(x_1, x_4, \E\right) = (0,0, Y > 140)$, i.e., individuals who exercise, do not have heart disease, and have high blood pressure, if they had not exercised previously and did not have heart disease, i.e., receiving treatment \((x_1, x_4) = (1,0)\), their blood pressure would slightly increase by 1.75.

 Since our data generation mechanism ensures the exact same observed data distribution as described in \citet{lu2023evaluating} after binarizing $Y$, we directly use their posterior total causal effects in their  Table 1 for comparative analysis. Specifically, we will adopt the definitions of the binary outcomes from \citet{lu2023evaluating}, denoted as $\mathrm{PN}^*(X_k\Rightarrow \E)$ and $\operatorname{postTCE}^*(X_k \Rightarrow \E \mid x, Y>140)$, as follows: 
\begin{gather*}
    \mathrm{PN}^*(X_k\Rightarrow \E)=\pr(Y_{X_k=0}<140\mid a_k,X_k=1,d_k,Y  >140),\\
    \operatorname{postTCE}^*(X_k \Rightarrow  \E \mid  x, Y>140)=\pr(Y_{X_k=1}>140\mid x,Y>140)-\pr(Y_{X_k=0}>140\mid x,Y>140).
\end{gather*} 
We use the symbol asterisks (*) to differentiate from the definitions provided in this paper. 
The definitions mentioned above can be identified using Lemma 1 and Theorem 1 in \citet{lu2023evaluating}.   We do not consider the causal estimands  $\operatorname{postNDE}^*(X_k \Rightarrow  \E \mid  x, Y>140)$ and $\operatorname{postNIE}^*(X_k \Rightarrow  \E \mid  x, Y>140)$, because there is no literature to support the identifiability results for these two causal quantities after binarization.

\begin{table}[t]
    \centering
    \caption{Results of marginal probabilities of necessity and posterior causal estimands based on the evidence $\{X=(1,1,1,1,1), Y>140\}$.}
    \label{tab:evidence-post-11111}
    \begin{threeparttable}
\resizebox{0.804918\columnwidth}{!}{% 
        \begin{tabular}{ccccccc} \toprule
            &&  $X_1$ & $X_2$ & $X_3$ & $X_4$ & $X_5$ \\\cline{3-7}\addlinespace[1mm]
            $\mathrm{{PN}}^*(X_k\Rightarrow \E) $& & 0.347 & 0.230 & 0.133 & 0.760 & 0.563 \\
            $\operatorname{postTCE}^*(X_k \Rightarrow  \E \mid  x, Y>140)$ && 0.344 & 0.207 & 0 & 0.722 & 0 \\
             $\operatorname{postNDE}(X_k \Rightarrow  Y \mid  x, Y>140)$ &&  3.823 & 0 & 0 & 17.023 & 0 \\  
             $\operatorname{postNIE}(X_k \Rightarrow  Y \mid  x, Y>140)$ &&6.805 & 4.561 & 0 & 0 & 0 \\  
             $\operatorname{postTCE}(X_k \Rightarrow  Y \mid  x, Y>140)$ & &10.628 & 4.561 & 0 & 17.023 & 0  \\\bottomrule
        \end{tabular} }
    \end{threeparttable}
\end{table}
 Given the observed evidence $\{X=(1,1,1,1,1),Y>140\}$, Table \ref{tab:evidence-post-11111} presents the results for each potential risk factor $X_k$ with respect to  posterior causal estimands. The first row of Table \ref{tab:evidence-post-11111} displays the marginal probabilities of necessity $\mathrm{PN}^*(X_k\Rightarrow \E)$ computed after binarizing the outcome variable $Y$, while the second row presents the posterior total causal effects $\operatorname{postTCE}^*(X_k \Rightarrow \E \mid x, Y>140)$. 
 The third to fifth rows present the results for posterior natural direct effects, indirect effects, and total causal effects considered in this paper.  
We find that the second and fifth rows of the table show very similar results in terms of sign and ordering. Specifically, HD, E, and D (denoted as $X_1$, $X_2$, and $X_4$, respectively) all have non-zero postTCE values, indicating that they are risk factors for blood pressure. Among them, HD  (i.e., $X_4$) has the largest postTCE, indicating that HD is the most important risk factor for blood pressure. In addition, the value of $\mathrm{PN}^*(X_4 \Rightarrow \E)$ was also the largest, further confirming the importance of HD as a risk factor. Notably, for HD  (i.e., $X_4$), it can be observed that the postNDE is equal to the postTCE of BP. On the other hand, the postNDE for E (i.e., $X_1$) is smaller than the postTCE for BP, implying that E has direct and indirect causal effects on BP. 
The third to fifth rows of the table indicate the postNDE, postNIE, and postTCE values for Hb and CP (denoted as $X_3$ and $X_5$, respectively), which are all equal to zero. This indicates that they are not risk factors for blood pressure. From the causal network in Figure \ref{fig:BP-DAG},   it can be observed that there is no causal path from Hb (i.e., $X_3$)  and CP (i.e., $X_4$)  to BP.  However, $\mathrm{PN}^*(X_5 \Rightarrow \E)$ is greater than $\mathrm{PN}^*(X_1 \Rightarrow \E)$ and $\mathrm{PN}^*(X_2 \Rightarrow \E)$ as shown in the first row.

\begin{table}[t]
    \centering 
    \caption{Results of marginal probabilities of necessity and posterior causal estimands based on the evidence $\{X=(1,1,1,0,1), Y>140\}$.}
    \label{tab:evidence-post-11101} 
    \resizebox{0.9\textwidth}{!}{%
        \begin{threeparttable}
            \begin{tabular}{ccccccc} 
                \toprule
                & & $X_1$ & $X_2$ & $X_3$ & $X_4$ & $X_5$ \\
                \cline{3-7}
                $\mathrm{{PN}}^*(X_k\Rightarrow \E)$ & & 0.347 & 0.230 & 0.133 & $\#$ & 0.563 \\ 
                $\operatorname{postTCE}^*\left(X_k \Rightarrow \E \mid  x, Y>140\right)$ & & 0.200 & 0 & 0 & 0 & 0  \\
                $\operatorname{postNDE}\left(X_k \Rightarrow Y \mid  x, Y>140\right)$ & & 2.000 & 0 & 0 & 10.483 & 0 \\ 
                $\operatorname{postNIE}\left(X_k \Rightarrow Y \mid  x, Y>140\right)$ & & 0 & 0 & 0 & 0 & 0 \\ 
                $\operatorname{postTCE}\left(X_k \Rightarrow Y \mid  x, Y>140\right)$ & & 2.000 & 0 & 0 & 10.483 & 0 \\ 
                \bottomrule
            \end{tabular}  
            \begin{tablenotes}
                \item[]\# Corresponding quantity is undefined.
            \end{tablenotes}
        \end{threeparttable}
    }
\end{table}

Given the observed evidence $\{X=(1,1,1,0,1),Y>140\}$, Table \ref{tab:evidence-post-11101} presents the results for each possible risk factor $X_k$ with respect to  posterior causal estimands. The first two rows of results are directly obtained from  \citet{lu2023evaluating}  for comparison purposes. In the second row, we observe that $\operatorname{postTCE}^*\left(X_1 \Rightarrow \E \mid \cdot\right) =0.2$, and $\operatorname{postTCE}^*\left(X_k \Rightarrow \E \mid \cdot\right) =0$ for $k=2, \ldots, 5$. Our results regarding postTCE are presented in the fifth row, showing a significant increase in blood pressure by 10.483 units when suffering from heart disease, i.e., $\operatorname{postTCE}^*\left(X_4 \Rightarrow \E \mid \cdot\right) =10.483$. This is because \citet{lu2023evaluating}  additionally require the monotonicity assumption of the outcome variable for identifiability in the binary outcome case. This assumption ensures that given the evidence $X_ 4= 0$ (indicating no heart disease), $X_ 1$ is the unique risk factor. However, the perfect rank assumption introduced for the continuous variable does not guarantee this. It can also be observed that the probability of necessity is not zero for $X_2$, $X_3$, and $X_5$; $\operatorname{postTCE}^*\left(X_k \Rightarrow \E \mid \cdot\right)$, $\operatorname{postTCE}\left(X_k \Rightarrow Y \mid \cdot\right)$, and $\operatorname{postNDE}\left(X_k \Rightarrow Y \mid \cdot\right)$ are zero when $X_4=0$ for $k=1,2,5$. Additionally, it can be observed from the fourth row that $\operatorname{postNIE}\left(X_k\Rightarrow Y \mid \cdot\right)=0$ for all $k=1,\ldots,5$, suggesting that all causes have no indirect effect on BP. From the causal network in Figure \ref{fig:BP-DAG}, since the evidence indicates no heart disease, i.e., $X_4=0$, it can be intuitively understood that $X_4$ blocks other nodes from transmitting effects to the outcome along the paths.
\begin{table}[t]
    \centering
    
    \caption{Results of marginal probabilities of necessity and posterior causal estimands based on the evidence $\{X=(1,0,1,1,1), Y>140\}$.}
    \label{tab:evidence-post-10111}
    \begin{threeparttable}
\resizebox{0.74804918\columnwidth}{!}{% 
      \begin{tabular}{ccccccc}\toprule
&& $X_1$ & $X_2$ & $X_3$ & $X_4$ & $X_5$ \\\cline{3-7}\addlinespace[1mm]
$\operatorname{postTCE}^*\left(X_k \Rightarrow \E \mid  x, Y>140\right)$ && 0.449 & 0 & 0 & 0.722 & 0 \\ $\operatorname{postNDE}\left(X_k \Rightarrow Y\mid  x, Y>140\right)$  && 3.424 & 0& 0& 16.856 & 0\\ 
  $\operatorname{postNIE}\left(X_k \Rightarrow Y\mid  x, Y>140\right)$  & & 8.994 & 0& 0& 0& 0\\ 
  $\operatorname{postTCE}\left(X_k \Rightarrow Y\mid  x, Y>140\right)$ &  & 12.418 & 0& 0& 16.856 & 0\\ \bottomrule
\end{tabular} }
    \end{threeparttable}
\end{table}

For the observed evidence $\{X=(1,0,1,1,1), Y>140\}$, where $X_2=0$ indicates a healthy diet, the values of postTCE and postNDE are shown in Table \ref{tab:evidence-post-10111}. Comparing it with Table  \ref{tab:evidence-post-11111}, we see that
\begin{align*}
\begin{gathered}
\operatorname{postTCE}\left\{X_1 \Rightarrow Y \mid X=(1,0,1,1,1), Y>140\right\}=12.418 \\
~~>\operatorname{postTCE}\left\{X_1 \Rightarrow Y \mid X=(1,1,1,1,1), Y>140\right\}=10.628,
\end{gathered}
\end{align*} 
and
\begin{align*}
\begin{gathered}
\operatorname{postTCE}\left\{X_2 \Rightarrow Y \mid X=(1,0,1,1,1), Y>140\right\}=0 \\
~~~~<\operatorname{postTCE}\left\{X_2 \Rightarrow Y \mid X=(1,1,1,1,1), Y>140\right\}=4.561.
\end{gathered}
\end{align*} 
This is because changing $X_2 = 1$ to $X_2 = 0$ in the evidence increases the attribution to $X_1$ and eliminates the attribution to $X_2$. Similar conclusions are also found in \citet{lu2023evaluating} in the binary case.  To assess the stability of the proposed estimation procedure in \ref{sec:estimation}, we conducted simulation studies by generating data according to the causal network depicted in Figure \ref{fig:BP-DAG}. The estimated results of Tables \ref{tab:evidence-post-11111}, \ref{tab:evidence-post-11101}, and \ref{tab:evidence-post-10111} under different sample sizes were obtained. The simulation results indicate negligible biases and small standard errors, particularly for large sample sizes. For detailed information on data generation procedures and simulation results, please refer to the supplementary material.

 \subsection{Example 2: NTP dataset} 
 \label{sec:ntp}
In this section, we apply the proposed method to a real dataset from the developmental toxicology experiments conducted by the National Toxicology Program (NTP)  \citep{ntp_tr602}. The primary objective of this study is to analyze whether tris(1-chloro-2-propyl) phosphate (TCPP) is a risk factor for abnormal weight loss in B6C3F1/N mice, or if there are other causes that are the potential risks.   In this experiment, a total of 120 mice were randomly exposed to six different dose levels of TCPP via dosed feed: 0, 1250, 2500, 5000, 10000, or 20000 ppm, for a duration of 3 months. Each pup's data includes gender (male/female), weekly body weights for three months, organ weights, and whether organs exhibit pathology. All data were completely collected without any missing values. In our analysis, let $X_1$ represent the gender of the mouse, where $X_{1}=0$ indicates female mice and $X_{1}=1$ indicates male mice. Let $X_{2}=0$ represent exposure to the low-dose group, including 0, 1250, and 2500 ppm, while $X_{2}=1$ represents exposure to the high-dose group, including 5000, 10000, or 20000 ppm. Let $X_{3}$ denote whether the mouse's liver or kidney exhibits pathology, where  $X_{3}=0$ indicates no pathology and $X_{3}=1$ indicates pathology. We choose the body weight at the end of three months as the outcome variable $Y$. In this analysis, we focus on assessing the potential risk factors affecting the body weight of underweight mice, indicated by the event $\E=\mathbb{I}(Y<27)$. 

\begin{figure}
    \centering
    \includegraphics[width=0.889175\textwidth]{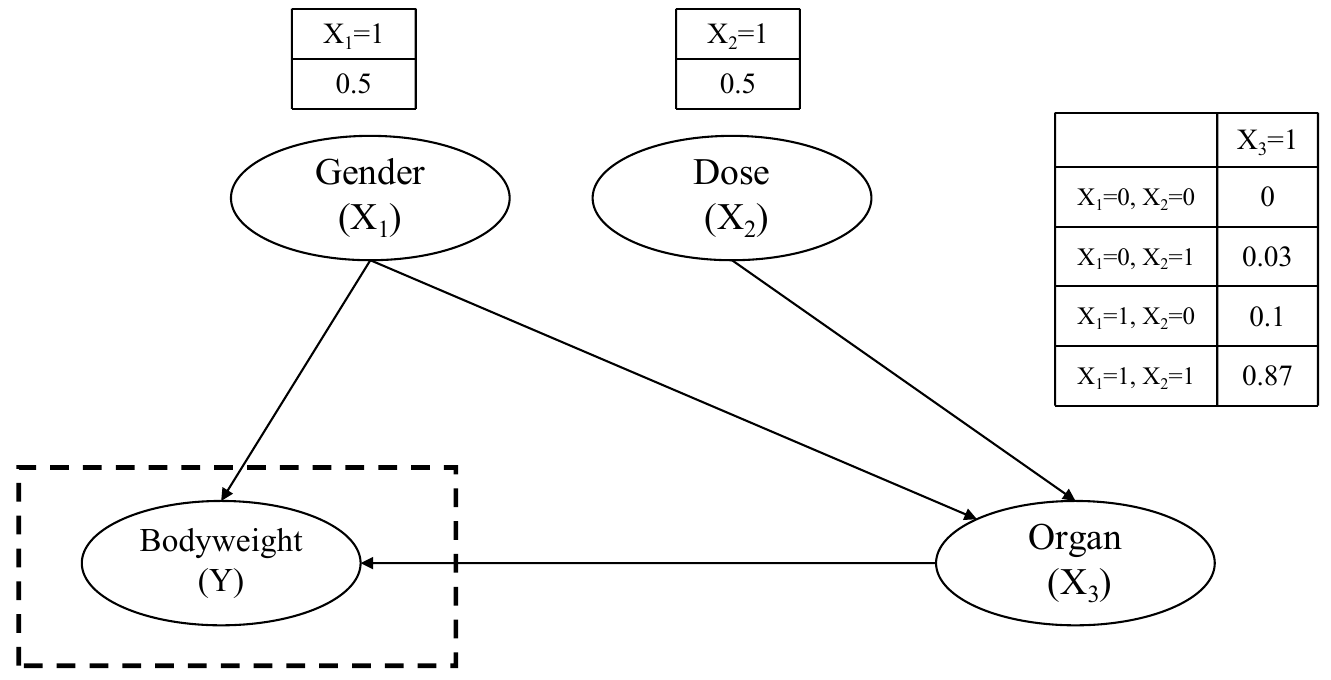}
    \caption{A causal network representing developmental toxicology experiments, including body weight and its potential risk factors.}
    \label{fig:DP-DAG}
\end{figure}

 We first use the R package ``{bnlearn}" to construct a Bayesian network based on the collected data as shown in Figure \ref{fig:DP-DAG}. It is clear that dose  $X_2$  affects body weight $Y$ indirectly through the occurrence of organ abnormalities $X_3$. Since both gender (i.e., \(X_1\)) and dose (i.e., \(X_2\)) can be considered as randomized trials, we can estimate the causal effect of \(X_1\) on \(Y\) using a mean difference estimator, i.e., \( E(Y_{X_1=1}-Y_{X_1=0})\) of 6.09; which implies that males are heavier. Similarly, we can estimate the causal effect of \(X_2\) on \(Y\) to be -3.353, which suggests that higher doses lead to weight loss.   Figure \ref{fig:DP-DAG} also provides a simple descriptive statistical analysis of these potential risk factors. We observe that as the toxin level increases, the occurrence rate of organ abnormalities also increases. In addition, males showed higher variability in organ abnormalities compared to females, as demonstrated in previous studies  \citep{bianco2023sex}.  These empirical findings align with monotonicity assumption \ref{assump:monotonicity}, i.e., $X_3(0,0)\leq \{ {X_3(0,1)},  {X_3(1,0)} \}\leq  {X_3(1,1)}$. In this data analysis, we choose not to binarize the outcome variable and therefore do not report the results of the binarized posterior causal effect, as the outcome variable fails to satisfy the monotonicity assumption regarding the causes (Assumption 2(b) in \citet{lu2023evaluating}).

% We initially present the posterior intervention causal effects for underweight based on various observed evidence in Table \ref{tab: NTP-res}. When considering the evidence $(X_1 ,X_3 ,\E )=(1,0,Y<27)$, we observe that changing from being a male mouse with no organ abnormalities to being a female mouse with organ abnormalities, i.e., $(X_1 ,X_3 )=(0,1)$, resulted in the most significant weight loss, totaling 10.51. For the observed evidence $(X_1 ,X_3 ,\E )=(0,0,Y<27)$, changing from a female mouse with organ abnormalities to one without led to a slight increase in weight, amounting to 1.42. In summary, under the given posterior evidences, transitioning from female $X_1=0$ to male mice $X_1=1$ increases weight, while transitioning from mice with no organ abnormalities $X_3=0 $   to those with organ abnormalities $X_3=1 $  decreases weight in most cases.

We now present the estimation results of posterior intervention causal effects based on various observed evidence in Table \ref{tab: NTP-res}.
 When considering the evidence $(X_1 ,X_3 ,\E )=(1,0,Y<27)$, we observed that changing from male mice without organ abnormalities to female mice with organ abnormalities (i.e., $(X_1 ,X_3 )=(0,1)$) led to the most significant body weight loss, which totaled 10.51. For the observed evidence $(X_1 ,X_3 ,\E )=(0,0,Y< 27)$, changing from female mice with organ abnormalities to female mice without organ abnormalities resulted in a slight increase in body weight of 1.42. In summary, for a given posterior evidence, changing from female mice $X_1=0$ to male mice $X_1=1$ resulted in an increase in body weight, while changing from mice without organ abnormalities $X_3=0$ to mice with organ abnormalities $X_3=1$ resulted in a decrease in body weight in most cases.

\begin{table}[t]
\centering
\caption{Results of posterior intervention causal effect based on different evidences.}
\label{tab: NTP-res}   
\resizebox{0.91848\columnwidth}{!}{% 
\begin{tabular}{ccccccc}
 \toprule   \addlinespace[0.5mm]\addlinespace[0.25mm]
    $ \mathrm{PostICE}(Y_{ x^{\prime}}   \mid  x, Y<27)$  & & $(x_1,x_3)=(0,0)$ & $(x_1,x_3)=(0,1)$ & $(x_1,x_3)=(1,0)$  &$(x_1,x_3)=(1,1)$  \\ 
 \addlinespace[0.25mm]\cline{3-6} \addlinespace[1mm] 
 $(x_1^\prime,x_3^\prime)=(0,0)$ &&  0.00 & 1.59 & -7.10 & -2.77 \\  \addlinespace[0.25mm]
 $(x_1^\prime,x_3^\prime)=(0,1)$  &&    -1.42 & 0.00 & -5.50 & -3.35 \\  \addlinespace[0.25mm]
 $(x_1^\prime,x_3^\prime)=(1,0)$
 & &8.81 & 10.51 & 0.00 & 5.58 \\ 
 $(x_1^\prime,x_3^\prime)=(1,1)$  &&2.84 & 4.46 & -8.90 & 0.00 \\ \addlinespace[0.25mm] \bottomrule 
\end{tabular} }
\end{table}

Table \ref{tab:NTP-data} presents the estimation results for each potential risk factor \(X_k\) with respect to posterior causal estimands.  Given the evidence $(X_1,X_2,X_3,\mathcal{E})=(1,1,1,Y<27)$, we find that the PostNDE value for gender \(X_1\) is the highest, indicating that gender might be the most important direct factor among the three considered factors affecting low weight, followed by organ abnormality \(X_3\). The PostNDE for toxin dosage \(X_2\) is 0, consistent with the conclusions drawn from the Bayesian network in Figure \ref{fig:DP-DAG}. Additionally, we observe that both gender \(X_1\) and toxin dosage \(X_2\) have indirect effects on the outcome variable, with the absolute value of the PostNIE for gender \(X_1\) significantly greater than toxin dosage \(X_2\), while organ abnormality  \(X_3\) has no indirect effect, suggesting that gender might be the most important indirect factor affecting low weight among the three factors, followed by toxin dosage. However, in terms of the posterior total causal effect, the influence of organ abnormality \(X_3\) is the greatest, indicating that organ abnormality is the most crucial risk factor leading to low weight, followed by toxin dosage \(X_2\), with the influence of gender \(X_1\) being the mildest.  
In summary, for most evidences in Table \ref{tab:NTP-data}, either gender \(X_1\) or organ abnormality \(X_3\) has the highest absolute PostTCE value, especially \(X_3\), suggesting that organ abnormality is the most important risk factor leading to weight loss in most cases. Meanwhile, under the evidences $(1,0,0,Y<27)$ and $(0,1,1,Y<27)$, toxin dosage \(X_2\) exhibits the most significant PostNIE value for weight loss, indicating that toxin dosage \(X_2\)  is the most important indirect risk factor in these situations.
\begin{table}[t]
\centering
\caption{Results of posterior causal estimands based on the various evidences for the NTP dataset.} 
\label{tab:NTP-data} 
\resizebox{0.997501\textwidth}{!}{%
\begin{tabular}{cccccccccc}
\toprule
                                                      &  & \multicolumn{8}{c}{Evidence $X=x $}                                                                                                                                                                                                                                                 \\ \cline{3-10}\addlinespace[1.3mm]
                                             Posterior    causal estimands    &  & \multicolumn{1}{c}{$( 0,0,0 ) $} & \multicolumn{1}{c}{$( 1,0,0 )$} & \multicolumn{1}{c}{$( 0,1,0 )$} & \multicolumn{1}{c}{$( 1,1,0 )$} & \multicolumn{1}{c}{$ ( 0,0,1 ) $} & \multicolumn{1}{c}{$ ( 1,0,1 ) $} & \multicolumn{1}{c}{$( 0,1,1 )$} & \multicolumn{1}{c}{$( 1,1,1 ) $} \\ \cline{1-1} \cline{3-10}\addlinespace[1.3mm]
$ \mathrm{PostNDE}(X_1 \Rightarrow Y \mid x, Y<27 ) $ &  & 8.83                             & 7.00                            & 8.83                            & 7.00                            & 10.00                             & 8.31                              & 10.00                           & 8.12                             \\
$ \mathrm{PostNDE}(X_2 \Rightarrow Y \mid x, Y<27 ) $ &  & 0.00                             & 0.00                            & 0.00                            & 0.00                            & 0.00                              & 0.00                              & 0.00                            & 0.00                             \\
$ \mathrm{PostNDE}(X_3 \Rightarrow Y \mid x, Y<27 ) $ &  & -1.43                            & -8.90                           & -1.43                           & -8.90                           & -6.30                             & -5.51                             & -6.30                           & -5.51                            \\\addlinespace[1mm]
$ \mathrm{PostNIE}(X_1 \Rightarrow Y \mid x, Y<27 ) $ &  & -0.60                            & 0.00                            & -5.18                           & 0.00                            & 0.00                              & -5.51                             & 0.00                            & -5.30                            \\
$ \mathrm{PostNIE}(X_2 \Rightarrow Y \mid x, Y<27 ) $ &  & -0.05                            & -7.58                           & 0.00                            & 0.00                            & 0.00                              & 0.00                              & -6.30                           & -4.87                            \\
$ \mathrm{PostNIE}(X_3 \Rightarrow Y \mid x, Y<27 ) $ &  & 0.00                             & 0.00                            & 0.00                            & 0.00                            & 0.00                              & 0.00                              & 0.00                            & 0.00                             \\\addlinespace[1mm]
$ \mathrm{PostTCE}(X_1 \Rightarrow Y \mid x, Y<27 ) $ &  & 8.23                             & 7.00                            & 3.65                            & 7.00                            & 10.00                             & 2.80                              & 10.00                           & 2.82                             \\
$ \mathrm{PostTCE}(X_2 \Rightarrow Y \mid x, Y<27 ) $ &  & -0.05                            & -7.58                           & 0.00                            & 0.00                            & 0.00                              & 0.00                              & -6.30                           & -4.87                            \\
$ \mathrm{PostTCE}(X_3 \Rightarrow Y \mid x, Y<27 ) $ &  & -1.43                            & -8.90                           & -1.43                           & -8.90                           & -6.30                             & -5.51                             & -6.30                           & -5.51                            \\ \bottomrule
\end{tabular}}
\end{table}
\section{Discussion} 
 
 \citet{dawid2014fitting}   pointed out that statistical inference about causes of effects for a specified individual is particularly problematic from many points of view and is difficult to justify even in ideal circumstances. While some prior research has addressed attribution analysis problems with binary outcome variables \citep{pearl2000,pearl2015causes,dawid2014fitting,dawid2015causes,dawid2022bounding,Li2023BKA,lu2023evaluating}, evaluating causes of continuous outcomes or effects  remains underexplored.  In this paper, we conduct attribution analysis for continuous outcomes within the framework proposed by  \citet{lu2023evaluating}.  Specifically, we propose a series of posterior causal estimands for retrospectively evaluating multiple correlated causes from a continuous effect. We also propose an efficient estimation procedure and establish corresponding asymptotic properties.
 
 The identifiability of individual treatment  effects and posterior intervention causal effects with multiple potentially correlated causes is motivated by the identification of individual treatment  effects in the case of a single cause \citep{heckman1997making}.   However, even if individual treatment effects are identified, it is not sufficient for retrospective assessment of a specific cause of effects, as the monotonicity assumption \ref{assump:monotonicity} is still necessary.   In practice, for estimation purposes, we can begin by recovering the counterfactual mappings between different potential outcomes as well as individual treatment effects through quantile regression \citep{koenker1978regression,heckman1997making,feng2020estimation}; then, we can utilize the identification expressions in Lemma \ref{lem:ice} and Theorem \ref{thm:medeff} for nonparametric estimation. 

% In addition to continuous outcomes, the attribution analysis for continuous causes is common in practice and is a matter of great interest. Besides, the problem of retrospective analysis when multiple causes and multiple continuous variables are involved in many medical diagnostic problems is also worth exploring \citep{Li2023BKA}. Finally, it is also interesting to discuss the bounds of the proposed estimates when the monotonicity assumption does not hold. However, these issues are beyond the scope of this paper and are left for future research.

In addition to continuous outcomes,  the attribution analysis of continuous causes is common in practice and is an issue of great interest. Second, the problem of retrospective analysis when multiple causes and multiple continuous outcomes are involved in many medical diagnoses is also worth exploring \citep{Li2023BKA}. Finally, it is also interesting to discuss the bounds of the proposed estimands or to consider sensitivity analysis when the monotonicity assumption does not hold \citep{tian2000probabilities,dawid2022bounding}. However, these issues are beyond the scope of this paper and are left for future research.
\section*{Supplementary material}
\label{SM}
The supplementary material contains proofs of all theoretical results, identiﬁability results under a causal network, and simulation details  for evaluation of the proposed procedure.
 
 \bibliographystyle{apalike}
					\bibliography{mybib}

\begin{thebibliography}{}

\bibitem[Bianco et~al., 2023]{bianco2023sex}
Bianco, A., Antonacci, Y., and Liguori, M. (2023).
\newblock Sex and gender differences in neurodegenerative diseases: Challenges
  for therapeutic opportunities.
\newblock {\em International Journal of Molecular Sciences}, 24(7):6354.

\bibitem[Chernozhukov and Hansen, 2005]{chernozhukov2005nonparametric}
Chernozhukov, V. and Hansen, C. (2005).
\newblock Nonparametric instrumental variable estimation under monotonicity.
\newblock {\em Econometrica}, 73(5):1351--1401.

\bibitem[Dawid et~al., 2014]{dawid2014fitting}
Dawid, A.~P., Faigman, D.~L., and Fienberg, S.~E. (2014).
\newblock Fitting science into legal contexts: Assessing effects of causes or
  causes of effects?
\newblock {\em Sociological Methods \& Research}, 43(3):359--390.

\bibitem[Dawid et~al., 2015]{dawid2015causes}
Dawid, A.~P., Faigman, D.~L., and Fienberg, S.~E. (2015).
\newblock On the causes of effects: Response to pearl.
\newblock {\em Sociological Methods \& Research}, 44(1):165--174.

\bibitem[Dawid et~al., 2024]{dawid2022bounding}
Dawid, P., Humphreys, M., and Musio, M. (2024).
\newblock Bounding causes of effects with mediators.
\newblock {\em Sociological Methods \& Research}, 53(1):28--56.

\bibitem[Feng et~al., 2020]{feng2020estimation}
Feng, Q., Vuong, Q., and Xu, H. (2020).
\newblock Estimation of heterogeneous individual treatment effects with
  endogenous treatments.
\newblock {\em Journal of the American Statistical Association},
  115(529):231--240.

\bibitem[Galhotra et~al., 2021]{galhotra2021explaining}
Galhotra, S., Pradhan, R., and Salimi, B. (2021).
\newblock Explaining black-box algorithms using probabilistic contrastive
  counterfactuals.
\newblock In {\em Proceedings of the 2021 International Conference on
  Management of Data}, pages 577--590.

\bibitem[Heckman et~al., 1997]{heckman1997making}
Heckman, J.~J., Smith, J., and Clements, N. (1997).
\newblock Making the most out of programme evaluations and social experiments:
  Accounting for heterogeneity in programme impacts.
\newblock {\em The Review of Economic Studies}, 64(4):487--535.

\bibitem[Horvitz and Thompson, 1952]{horvitz1952generalization}
Horvitz, D.~G. and Thompson, D.~J. (1952).
\newblock A generalization of sampling without replacement from a finite
  universe.
\newblock {\em Journal of the American Statistical Association},
  47(260):663--685.

\bibitem[Imai et~al., 2010]{Imai2010Stasci}
Imai, K., Keele, L., and Yamamoto, T. (2010).
\newblock {Identification, Inference and Sensitivity Analysis for Causal
  Mediation Effects}.
\newblock {\em Statistical Science}, 25(1):51 -- 71.

\bibitem[Khoury et~al., 2004]{khoury2004epidemiologic}
Khoury, M.~J., Yang, Q., Gwinn, M., Little, J., and Flanders, W.~D. (2004).
\newblock An epidemiologic assessment of genomic profiling for measuring
  susceptibility to common diseases and targeting interventions.
\newblock {\em Genetics in Medicine}, 6(1):38--47.

\bibitem[Koenker and Bassett, 1978]{koenker1978regression}
Koenker, R. and Bassett, G. (1978).
\newblock Regression quantiles.
\newblock {\em Econometrica}, 46(1):33.

\bibitem[Korostelev and Tsybakov, 2012]{korostelev2012minimax}
Korostelev, A.~P. and Tsybakov, A.~B. (2012).
\newblock {\em Minimax theory of image reconstruction}, volume~82.
\newblock Springer Science \& Business Media.

\bibitem[Li et~al., 2023]{Li2023BKA}
Li, W., Lu, Z., Jia, J., Xie, M., and Geng, Z. (2023).
\newblock {Retrospective causal inference with multiple effect variables}.
\newblock doi: \url{https://doi.org/10.1093/biomet/asad056}.

\bibitem[Lu et~al., 2023]{lu2023evaluating}
Lu, Z., Geng, Z., Li, W., Zhu, S., and Jia, J. (2023).
\newblock Evaluating causes of effects by posterior effects of causes.
\newblock {\em Biometrika}, 110(2):449--465.

\bibitem[NTP, 2023]{ntp_tr602}
NTP (2023).
\newblock {NTP. TR-602: Tris(Chloropropyl)phosphate (13674-84-5). Chemical
  Effects in Biological Systems (CEBS). Research Triangle Park, NC (USA):
  National Toxicology Program (NTP).}
\newblock Accessed 2024-03-06.
  \url{https://cebs.niehs.nih.gov/cebs/publication/TR-602}.

\bibitem[Pearl, 1999]{pearl1999probabilities}
Pearl, J. (1999).
\newblock Probabilities of causation: Three counterfactual interpretations and
  their identification.
\newblock {\em Synthese}, 1(121):93--149.

\bibitem[Pearl, 2000]{pearl2000}
Pearl, J. (2000).
\newblock {\em Causality: Models, reasoning, and inference}.
\newblock Cambridge:Cambridge University Press.

\bibitem[Pearl, 2015]{pearl2015causes}
Pearl, J. (2015).
\newblock Causes of effects and effects of causes.
\newblock {\em Sociological Methods \& Research}, 44(1):149--164.

\bibitem[Robins, 2000]{robins2000marginal}
Robins, J.~M. (2000).
\newblock Marginal structural models versus structural nested models as tools
  for causal inference.
\newblock In {\em Statistical models in epidemiology, the environment, and
  clinical trials}, pages 95--133. Springer New York, New York, NY.

\bibitem[Rosenbaum and Rubin, 1983]{rosenbaum1983central}
Rosenbaum, P. and Rubin, D. (1983).
\newblock The central role of the propensity score in observational studies for
  causal effects.
\newblock {\em Biometrika}, 70(1):41--55.

\bibitem[Sanders et~al., 2021]{sanders2021differential}
Sanders, J., Faigman, D.~L., Imrey, P.~B., and Dawid, P. (2021).
\newblock Differential etiology: inferring specific causation in the law from
  group data in science.
\newblock {\em Ariz. L. Rev.}, 63:851.

\bibitem[Tian and Pearl, 2000]{tian2000probabilities}
Tian, J. and Pearl, J. (2000).
\newblock Probabilities of causation: Bounds and identification.
\newblock {\em Annals of Mathematics and Artificial Intelligence},
  28(1-4):287--313.

\bibitem[Van Der~Vaart et~al., 1996]{van1996weak}
Van Der~Vaart, A.~W., Wellner, J.~A., van~der Vaart, A.~W., and Wellner, J.~A.
  (1996).
\newblock {\em Weak Convergence}.
\newblock Springer.

\bibitem[VanderWeele, 2012]{vanderweele2012sufficient}
VanderWeele, T.~J. (2012).
\newblock {\em The Sufficient Cause Framework in Statistics, Philosophy and the
  Biomedical and Social Sciences}.
\newblock John Wiley \& Sons, Ltd.

\bibitem[Vuong and Xu, 2017]{vuong2017counterfactual}
Vuong, Q. and Xu, H. (2017).
\newblock Counterfactual mapping and individual treatment effects in
  nonseparable models with binary endogeneity.
\newblock {\em Quantitative Economics}, 8(2):589--610.

\end{thebibliography}
  
 \end{document}